\documentclass[10pt,twocolumn,twoside,journal]{IEEEtran}
\IEEEoverridecommandlockouts
\usepackage{subfigure} 
\usepackage{graphicx}

\usepackage{amsmath,graphicx,amssymb,mathtools,bm}
\usepackage{subfigure}
\usepackage{hyperref}
\usepackage{cite}
\usepackage{amsmath,amssymb,amsfonts}
\usepackage{algorithmic}
\usepackage{graphicx}
\usepackage{textcomp}
\usepackage{xcolor}
\usepackage{verbatim}  
\usepackage{graphicx}  
\usepackage{bm}  
\usepackage{mathrsfs} 
\usepackage{algorithm} 
\usepackage{algorithmic} 
\usepackage{booktabs}
\usepackage{textcomp}  
\usepackage{multirow}  
\usepackage{lettrine}   
\usepackage{graphicx}  
\usepackage{color}  
\usepackage{amsmath}
\usepackage{amssymb}
\usepackage{stfloats}
\usepackage{caption}
\usepackage{subfigure}

\usepackage{color}  
\usepackage{hyperref}
\hypersetup{
   colorlinks=true,
linkcolor=blue,
filecolor=magenta,      
urlcolor=cyan,
pdfborderstyle={/S/U/W 1} 
}
\UseRawInputEncoding

\def\BibTeX{{\rm B\kern-.05em{\sc i\kern-.025em b}\kern-.08em
    T\kern-.1667em\lower.7ex\hbox{E}\kern-.125emX}}

 \newtheorem{Proposition}{\bf Proposition}[section]
  
\newtheorem{remark}[Proposition]{Remark}
\newtheorem{Corollary}[Proposition]{Corollary}
\newenvironment{Proof}{{\indent \it Proof:}}{\hfill $\blacksquare$\par}
 
\DeclareMathSizes{12}{10.3}{7}{5}
\begin{document}
	\newcommand{\tabincell}[2]{\begin{tabular}{@{}#1@{}}#2\end{tabular}}


\title{Performance Bounds for Near-Field Localization with  Widely-Spaced Multi-Subarray \\
	 mmWave/THz MIMO
}

\author{Songjie Yang, Xinyi Chen, Yue Xiu, Wanting Lyu, \\
	 Zhongpei Zhang, \IEEEmembership{Member,~IEEE} and Chau Yuen, \IEEEmembership{Fellow,~IEEE}

\thanks{Songjie Yang, Xinyi Chen, Yue Xiu, Wanting Lyu,
	and Zhongpei Zhang are with the National Key Laboratory of Wireless Communications, University of Electronic Science and Technology of China, Chengdu 611731, China. 
	(e-mail:
	yangsongjie@std.uestc.edu.cn;chenxy@163.com; xiuyue12345678@163.com;
	lyuwanting@yeah.net;
	zhangzp@uestc.edu.cn). 
	Chau Yuen is with the School of Electrical and Electronics Engineering, Nanyang Technological University (e-mail: chau.yuen@ntu.edu.sg).}}
\maketitle

\begin{abstract}
	This paper investigates the potential of near-field localization using widely-spaced multi-subarrays (WSMSs) and analyzing the corresponding angle and range Cram\'er-Rao bounds (CRBs).  By employing the Riemann sum, closed-form CRB expressions are derived for the spherical wavefront-based WSMS (SW-WSMS).  We find that the CRBs can be characterized by the angular span formed by the line connecting the array's two ends to the target, and the different WSMSs with same angular spans but different number of subarrays have identical normalized CRBs. We provide a theoretical proof that, in certain scenarios, the CRB of WSMSs is smaller than that of uniform arrays. We further yield the closed-form CRBs for the hybrid spherical and planar wavefront-based WSMS (HSPW-WSMS), and its components can be seen as decompositions of the parameters from the CRBs for the SW-WSMS.  Simulations are conducted to validate the accuracy of the derived closed-form CRBs and provide further insights into various system characteristics. Basically, this paper underscores the high resolution of utilizing WSMS for localization, reinforces the validity of adopting the HSPW assumption, and, considering its applications in communications, indicates a promising outlook for integrated sensing and communications based on HSPW-WSMSs.
\end{abstract}
\begin{IEEEkeywords}
Widely-spaced multi-subarrays, closed-form Cram\'er-Rao bounds, spherical wavefront, hybrid spherical and planar wavefront.
\end{IEEEkeywords} 
\section{Introduction} 
The sixth generation (6G) will seamlessly integrate sensing and communication into a unified system, harnessing radio waves to perceive the physical world and create digital twins in the cyber realm. Networked sensing introduces a new realm of possibilities beyond mere communication, encompassing various applications such as device-based or device-free localization, imaging, environmental reconstruction and monitoring, as well as gesture and activity recognition. This expanded sensing capability brings forth additional performance dimensions to the International Mobile Telecommunications (IMT), including detection probability, sensing resolution, and accuracy in terms of range, velocity, and angles. The specific requirements of these dimensions may vary depending on application to application. For future localization and reconstruction applications, high sensing accuracy and resolution will be indispensable, while imaging applications will demand ultra-high resolution as the primary factor. In the context of gesture and activity recognition, the utmost priority lies in achieving optimal detection probability.

Wireless localization, the process of determining the geographical position of a mobile target or user in wireless networks, plays a vital role in numerous applications ranging from emergency services and asset tracking to location-based services. In recent years, the emergence of massive multiple input multiple output (MIMO) technology has introduced new opportunities and challenges in the field of wireless localization. The ability of massive MIMO to employ an excessive number of antennas at the base station offers the potential for significant improvements in localization accuracy, coverage, and capacity \cite{MIMO-LOC}.

As the array aperture futher increases to overcome the high attenuation of mmWave/THz propagation, the near-field effect, which breaks the planar wavefront assumption, should be taken seriously. This makes more challenging for near-field signal processing in extremely large (XL)-arrays. However, challenges often come with opportunities;  
XL-arrays bring new potentials for near-field localization that is capable of sensing the range without multi-frequency pilots.
In \cite{NL1,NL2,NL3,NL4}, near-field localization has been investigated with different methods, showing the potential of range estimation.
Although these methods are not discussing near-field localization in the context of XL-arrays, they can be applied for XL-array localization. With the XL-array depolyment, the authors in \cite{NCE1,NCE2,NCE3,NCE4,NCE5} have investigated how to estimate the angle and range for channel estimation.
In addition, the Cram\'er-Rao bound (CRB) provides a lower bound on the covariance matrix of any unbiased estimator of an unknown parameter, which is useful for understanding the fundamental limits of estimation accuracy and for evaluating the performance of different estimation algorithms. In \cite{CRB1}, the CRB was analyzed for three-dimensional near-field localization. Reference \cite{CRB2} explored the theoretical bounds on the accuracy of near-field localization in bi-static MIMO radar systems under the deterministic and stochastic models.
Particularly, the authors in \cite{CRB3} comprehensively discussed the CRB for mono-/bi-static localization with phased/MIMO array in the near-field region. 

On the other hand, the widely-spaced multi-subarray (WSMS) layout becomes promising for mmWave/THz array antennas \cite{WS1,WS2,WS3,WS4,WS5,WS-MIMO}, due to 1) array scalability and flexity, 2) manufacturing feasiblity, 3)
simplified circuitry and signal processing, and 4) size and weight considerations. 

By exploiting the spatial structure of
WSMSs, the hybrid spherical and planar wavefront (HSPW) assumption was adopted \cite{WS2,WS-MIMO}, where the planar wavefront (PW) and the spherical wavefront (SW) holds for the intra-subarray and the inter-subarray, respectively, for simplicity of singal processing. In this context, multi-subarray beamforming and capacity analysis have been investigated. Particularly,
with the same number of antennas, \cite{WS-MIMO} proved that WSMSs could provid stronger multiplexing cabilities 
than the uniform XL-arrays by increasing the inter-subarray spacing to enlarge the near-field effect. 

Essentially, the near-field effect not only benefits communications but also sensing, specifically enabling narrowband range estimation. To enhance range estimation, it is necessary to increase the array aperture to accommodate large Rayleigh distances. However, this entails a significant number of antennas for uniform arrays (UAs). Addressing this concern, the WSMS presents itself as a potential solution. Consequently, the following questions arise: 1) How does near-field localization performance change when employing the WSMS with varied inter-subarray spacing? Additionally, 2) under what circumstances is it relatively appropriate to utilize the HSPW assumption for near-field localization, as mentioned earlier in our statement regarding the SW model?

Last but not least, 
integrated sensing and communication (ISAC), promising for 6G wireless networks, revolutionizes connectivity by seamlessly combining sensing capabilities with advanced communication technologies, enabling devices to communicate, perceive, and interpret their surroundings in a highly intelligent and context-aware network. As previously mentioned, both communication and sensing applications share the need to enhance the near-field effect. Hence, studying near-field ISAC holds significant value.

Motivated by the above, we aim to investigate the potential of near-field localization with WSMSs by analyzing its angle/range CRB, and hope to build a bridge for near-field communication and sensing with the promising array layout WSMS.  The main contributions are as follows:\footnote{The source code of this paper is open in \url{ https://github.com/YyangSJ/Near-field-CRB} for readers studying.}

\begin{itemize}
	\item We employ a bi-static MIMO sensing system, where the hybrid beamforming architecture with a large number of antennas for the transmitter (TX), and the fully-digital beamforming architecture with a small number of antennas for the RX. In this sense, using the sine rule, the receiver (RX) parameters (angle-of-arrival and range) can be expressed by the TX parameters. In this case, we can know how the distance between the TX and the RX impacts the CRB performance.
	 Then,
	we give the general CRB expressions, with respect to (w.r.t.) the angle-of-departure (AoD) and the range of TX-target, which depend on the received signal-to-noise ratio (SNR) and the array manifold functions (AMFs). 
	\item Under the general CRB expression, we first discuss
 the SW-based WSMS (SW-WSMS), corresponding to a complicated array manifold.
	To yield more insights, the closed-form CRBs are derived by calculating the manifold functions and the sum formulas with the Riemann sum. We find that a psysical index, the angular span, can be used to characterize the CRB. Based on this finding, the CRBs of two WSMSs, with the same angular spans but different number of subarrays and different inter-subarray spacing, are discussed, for finding that they have the same normalized Fisher matrix or normalized CRB as the WSMS. In fact, this is a unique conclusion existing with near-field effects. Moreover, we compare the CRBs of WSMS with the UA under the same array aperture and number of antennas.
	\item We derive
	the closed-form CRBs of HSPW-based WSMS (HSPW-WSMS), corresponding to a relative simple array manifold. 
 	Based on this, we
	compare the CRBs of the SW-WSMS and HSPW-WSMS. Besides, some corollaries regarding the SW-WSMS are also derived with HSPW-WSMS.
	Particularly, the asymptotic CRB for the HSPW-WSMS is analyzed.
\end{itemize}

The rest of this paper is organized as follows: Section \ref{sys} discusses the bi-static MIMO sensing system with hybrid beamforming architectures and provides general CRB expressions. Sections \ref{CRB_SW} and \ref{CRB-HSPW} derive the closed-form CRBs for WSMSs based on the SW and the HSPW, respectively. In section \ref{SR}, several simulations are carried out to demonstrate our derivations and offer insights into various system characteristics. Finally, the summary and outlook are presented in Section \ref{Con}.

{\emph {Notations}}:
${\left(  \cdot  \right)}^{ *}$, ${\left(  \cdot  \right)}^{ T}$ and ${\left(  \cdot  \right)}^{ H}$ denote conjugate, transpose, conjugate transpose, respectively. $\Re\{\cdot\}$ is the real part sysmbol.  $\otimes$ denotes the Kronecker product. $\Vert\cdot\Vert_2$ and $\vert\cdot\vert$ represent the $\l_2$ norm and modulus, respectively.
 Finally, $\mathcal{CN}(\mathbf{a},\mathbf{A})$ is the complex Gaussian distribution with mean $\mathbf{a}$ and covariance matrix $\mathbf{A}$.
\section{System Model and CRB}\label{sys}
\subsection{Signal Model}\label{Signal}
 This paper considers a  bi-static MIMO sensing system, where the TX is equipped with $N_t$ antennas and $N_{\rm RF}$ RF chains using the hybrid beamforming architecture, while the RX is equipped with $N_r$ antennas using the fully-digital beamforming architecture.
The received training signals in an arbitrary frame have a form of\footnote{Here, only the line-of-sight path is considered for sensing.}
\begin{equation}\label{Y}
	\mathbf{y}=\alpha\sqrt{N_rN_t} \mathbf{g}_r \mathbf{g}_t^H \mathbf{F}\mathbf{s}+\mathbf{n},
\end{equation}
where $\alpha$ denotes the complex reflection coefficient or the path gain, $\mathbf{g}_r\in\mathbb{C}^{N_r\times 1}$ and $\mathbf{g}_t\in\mathbb{C}^{N_t\times 1}$ are the array manifolds of the RX and the TX, respectively, which depend on the specific array layout. $\mathbf{F}\triangleq\mathbf{F}_{\rm RF}\mathbf{F}_{\rm BB}\in\mathbb{C}^{N_t\times N_{ RF}}$ is the hybrid precoder with the analog precoder $\mathbf{F}_{\rm RF}$ and the baseband precoder $\mathbf{F}_{\rm BB}$, and $\mathbf{s}\in\mathbb{C}^{N_{ RF}\times 1}$ is the transmitted symbol. In this paper, we assume identical pilot symbols such that $\mathbf{S}\triangleq \sigma_p\mathbf{I}_{N_{ RF}}$, where $\sigma_p$ is the transmit power which is set to $1$ in this study.  Moreover,
 $\mathbf{n}\in\mathbb{C}^{N_r\times 1}$ is the noise matrix with each element following $\mathcal{CN}(0,\sigma_n^2)$.  

According to Eqn. (\ref{Y}), we define the received training signal matrix in the $t$-th frame ($t=1,\cdots,T$) by $\mathbf{Y}_t\triangleq\alpha\sqrt{N_rN_t} \mathbf{g}_r \mathbf{g}_t^H \mathbf{f}_t +\mathbf{N}_t$, where $\mathbf{f}_t\triangleq\mathbf{F}_t\mathbf{s}_t$. By collecting the $T$ training signals with $\widetilde{\mathbf{Y}}=[\mathbf{y}_1,\cdots,\mathbf{y}_T]\in\mathbb{C}^{N_t\times T}$, we have
\begin{equation}
\widetilde{\mathbf{Y}}=\alpha\sqrt{N_rN_t} \mathbf{g}_r \mathbf{g}_t^H \widetilde{\mathbf{F}}+\widetilde{\mathbf{N}},
\end{equation}
where $\widetilde{\mathbf{F}}\triangleq\left[\mathbf{f}_1,\cdots,\mathbf{f}_{T}\right]\in\mathbb{C}^{N_t\times T}$ and ${\mathbf{N}}\triangleq\left[\mathbf{n}_1,\cdots,\mathbf{n}_T\right]\in\mathbb{C}^{N_r\times T}$. 
 
 Vectorizing $\widetilde{\mathbf{Y}}$ yields
\begin{equation}
	\widetilde{\mathbf{y}}\triangleq{\rm vec}\left(\widetilde{\mathbf{Y}}\right)=\alpha\sqrt{N_rN_t} \left(\widetilde{\mathbf{F}}^T\mathbf{g}^*_t\right)\otimes \mathbf{g}_r+\widetilde{\mathbf{n}},
\end{equation}
where $\widetilde{\mathbf{n}}\triangleq{\rm vec}\left(\widetilde{\mathbf{N}}\right)$.
 
 \subsection{General CRB Expressions}
Consider 
$\mathbf{h}\triangleq\alpha\sqrt{N_rN_t} \left(\widetilde{\mathbf{F}}^T\mathbf{g}^*_t\right)\otimes \mathbf{g}_r$, the Fisher matrix w.r.t. $\bm{\xi}\in\mathbb{C}^{L\times 1}$, with $L$ being the number of parameters, is given by \cite{CRB4}
\begin{equation}
	\bm{\mathcal{F}}=\frac{2}{\sigma_n^2}\Re\left\{\left(\frac{\partial \mathbf{h}}{\partial \bm{\xi}}\right)\left(\frac{\partial \mathbf{h}}{\partial \bm{\xi}}\right)^H
	\right\}.
\end{equation}
Then, the CRB of the $l$-th parameter in $\bm{\xi}$ is 
\begin{equation}
	\textbf{CRB}_l=\left[\bm{\mathcal{F}}^{-1}\right]_{l,l}.
\end{equation}

\begin{figure}
	\centering 
	\includegraphics[height=5.5cm,width=8cm]{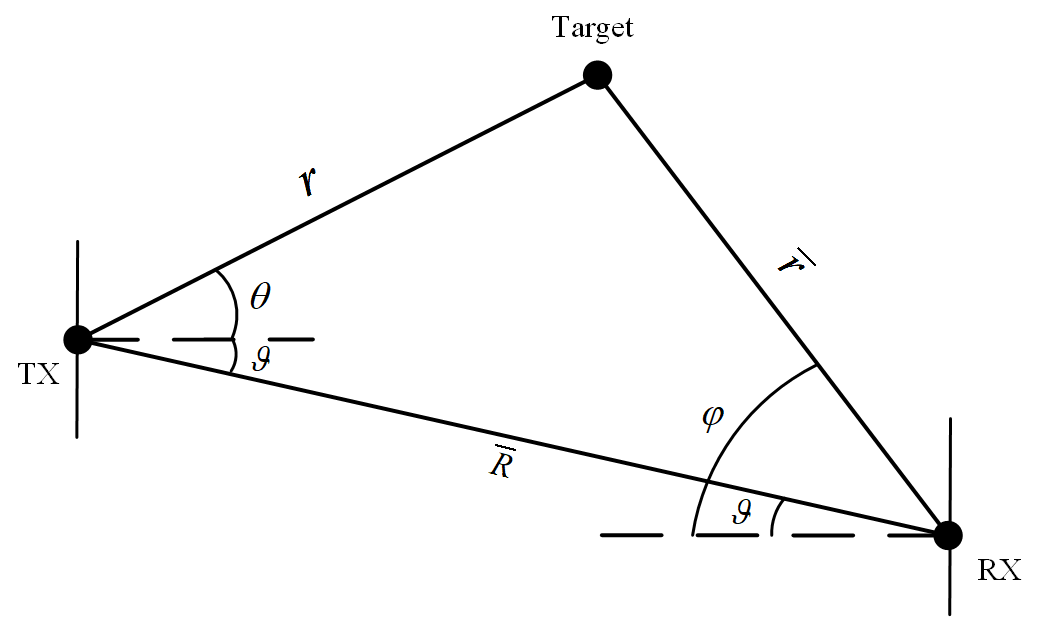}
	\caption{The bi-static sensing system.}\label{TX-RX} 
\end{figure} 
As shown in Fig. \ref{TX-RX}, we denote the AoD, the AoA, the relative angle, the distance between the TX and the target, and the distance between the TX and the RX by $\theta$, $\phi$, $\vartheta$, $r$, and $R$, respectively. The distance between the target and the RX can be calculated by $\overline{r}=\sqrt{R^2+r^2-2 R r \cos(\theta+\vartheta)}$. According to the sine rule such that $\frac{\sin (\phi-\vartheta)}{r}=\frac{\sin (\theta+\vartheta)}{\overline{r}}$, $\phi$ is expressed by
\begin{equation}\label{phiR}
	\phi(r, \theta)=\arcsin \left\{\frac{r \sin (\theta+\vartheta)}{\sqrt{R^2+r^2-2 R r \cos (\theta+\vartheta)}}\right\}+\vartheta.
\end{equation}  

Since we just consider the AoD $\theta$, the distance $r$, and the path gain $\alpha$, we set $\vartheta=0$ for clarity through this paper.
Therefore, $\bm{\xi}$ can be defined by $\bm{\xi}\triangleq[\theta,r,\alpha_R,\alpha_I]^T$, where $\alpha_R$ and $\alpha_I$ are the real and imaginary parts of $\alpha$, respectively. Denoted by $\beta\triangleq\alpha \sqrt{N_rN_t}$ and
\begin{equation}\label{wh}
\widetilde{\mathbf{h}}\triangleq\left(\widetilde{\mathbf{F}}^T\mathbf{g}^*_t(r,\theta)\right)\otimes \mathbf{g}_r(r,\theta),
\end{equation} 
we futher express $\bm{\mathcal{F}}$ as
\begin{equation}
	\bm{\mathcal{F}}=\frac{2}{\sigma_n^2}\begin{bmatrix}
		\bm{\Pi}_{1,1}&\bm{\Pi}_{1,2}\\ \bm{\Pi}_{1,2}^T&\bm{\Pi}_{2,2}
	\end{bmatrix},
\end{equation}
where $\bm{\Pi}_{1,1}\triangleq\begin{bmatrix}
	h_{\theta\theta} & h_{\theta r}\\h_{\theta r} & h_{r r}
\end{bmatrix}$,$\bm{\Pi}_{1,2}\triangleq\begin{bmatrix}
	h_{\theta\alpha_R} & h_{\theta \alpha_I}\\h_{ r\alpha_R} & h_{r \alpha_I}
\end{bmatrix}$,  $\bm{\Pi}_{2,2}\triangleq\begin{bmatrix}
	h_{\alpha_R\alpha_R} & h_{\alpha_R \alpha_I}\\h_{ \alpha_R\alpha_I} & h_{\alpha_I \alpha_I}
\end{bmatrix}$, and
$h_{l_1l_2}\triangleq \Re\left\{\left(\frac{\partial \mathbf{h}}{\partial \bm{\xi}_{l_1}}\right)^H\frac{\partial \mathbf{h}}{\partial \bm{\xi}_{l_2}}
\right\}$, $l_1,l_2\in\{1,\cdots,L\}$. Specifically, $h_{\theta\theta}\triangleq \Re\left\{\left(\frac{\partial \mathbf{h}}{\partial \theta}\right)^H\frac{\partial \mathbf{h}}{\partial \theta}
\right\}=\vert\beta\vert^2\left\Vert\frac{\partial \widetilde{\mathbf{h}}}{\partial \theta}\right\Vert_2^2$, $h_{\theta r}=\vert\beta\vert^2\Re\left\{\left(\frac{\partial \widetilde{\mathbf{h}}}{\partial \theta}\right)^H\frac{\partial \widetilde{\mathbf{h}}}{\partial r}
\right\}$, $h_{r r}=\vert\beta\vert^2\left\Vert\frac{\partial \widetilde{\mathbf{h}}}{\partial r}\right\Vert_2^2$, $h_{\theta \alpha_R}=\Re\left\{\beta^*\left(\frac{\partial \widetilde{\mathbf{h}}}{\partial \theta}\right)^H   \widetilde{\mathbf{h}} 
\right\}$, $h_{\theta \alpha_I}=\Re\left\{j\beta^*\left(\frac{\partial \widetilde{\mathbf{h}}}{\partial \theta}\right)^H   \widetilde{\mathbf{h}} 
\right\}=-\Im\left\{\beta^*\left(\frac{\partial \widetilde{\mathbf{h}}}{\partial \theta}\right)^H   \widetilde{\mathbf{h}} 
\right\}$, $h_{r \alpha_R}=\Re\left\{\beta^*\left(\frac{\partial \widetilde{\mathbf{h}}}{\partial r}\right)^H   \widetilde{\mathbf{h}} 
\right\}$, $h_{r \alpha_I}=\Re\left\{j\beta^*\left(\frac{\partial \widetilde{\mathbf{h}}}{\partial r}\right)^H   \widetilde{\mathbf{h}} 
\right\}=-\Im\left\{\beta^*\left(\frac{\partial \widetilde{\mathbf{h}}}{\partial r}\right)^H   \widetilde{\mathbf{h}} 
\right\}$, $h_{\alpha_R \alpha_R}=h_{\alpha_I \alpha_I}=\left\Vert\widetilde{\mathbf{h}}\right\Vert_2^2$, and $h_{\alpha_R \alpha_I}=\Re\left\{j\left\Vert\widetilde{\mathbf{h}}\right\Vert_2^2\right\}=0$. For clarity, $\widetilde{\mathbf{h}}_\theta\triangleq\frac{\partial \widetilde{\mathbf{h}}}{\partial \theta}$ and $\widetilde{\mathbf{h}}_r\triangleq\frac{\partial \widetilde{\mathbf{h}}}{\partial r}$ are defined through this paper.

We just focus on the CRB of $\theta$ and $r$, thus
\begin{equation}
	\bm{\mathcal{F}}^{-1}\triangleq\frac{\sigma_n^2}{2}\begin{bmatrix}
		\mathbf{Q}^{-1} &\bm{\times} \\ \bm{\times}&\bm{\times}
	\end{bmatrix},
\end{equation}
where $\mathbf{Q}\triangleq\bm{\Pi}_{1,1}-\bm{\Pi}_{1,2}\bm{\Pi}_{2,2}^{-1}\bm{\Pi}_{1,2}^T$ is the $2\times 2$ Schur complement, and $\bm{\times}$ denotes the element that we do not care.

Noticing that $\bm{\Pi}_{2,2}$ is a diagonal matrix due to $h_{\alpha_R\alpha_I}\triangleq\Re\left\{\left(\frac{\partial \mathbf{h}}{\partial \alpha_R}\right)^H\frac{\partial \mathbf{h}}{\partial \alpha_I}\right\}\equiv0$. Hence, $\mathbf{Q}$ is simplified as Eqn. (\ref{QQ}).
\begin{figure*}
\begin{equation}\label{QQ}
	\begin{aligned}
		\mathbf{Q}=&\begin{bmatrix}
			\vert\beta\vert^2\left\Vert\frac{\partial \widetilde{\mathbf{h}}}{\partial \theta}\right\Vert_2^2 & \vert\beta\vert^2\Re\left\{\left(\frac{\partial \widetilde{\mathbf{h}}}{\partial \theta}\right)^H\frac{\partial \widetilde{\mathbf{h}}}{\partial r}
			\right\}\\\vert\beta\vert^2\Re\left\{\left(\frac{\partial \widetilde{\mathbf{h}}}{\partial \theta}\right)^H\frac{\partial \widetilde{\mathbf{h}}}{\partial r}
			\right\} &\vert\beta\vert^2\left\Vert\frac{\partial \widetilde{\mathbf{h}}}{\partial r}\right\Vert_2^2
		\end{bmatrix}-\begin{bmatrix}
			\Re\left\{\beta^*\left(\frac{\partial \widetilde{\mathbf{h}}}{\partial \theta}\right)^H   \widetilde{\mathbf{h}} 
			\right\}& -\Im\left\{\beta^*\left(\frac{\partial \widetilde{\mathbf{h}}}{\partial \theta}\right)^H   \widetilde{\mathbf{h}}\right\} \\ 
			\Re\left\{\beta^*\left(\frac{\partial \widetilde{\mathbf{h}}}{\partial r}\right)^H   \widetilde{\mathbf{h}} 
			\right\} & -\Im\left\{\beta^*\left(\frac{\partial \widetilde{\mathbf{h}}}{\partial r}\right)^H   \widetilde{\mathbf{h}}
			\right\} 
		\end{bmatrix}\\ 
		& \ \   \ \ \ \ \ \ \ \ \ \ \ \ \ \ \ \ \ \ \  \ \times \begin{bmatrix}
			\frac{1}{\left\Vert\widetilde{\mathbf{h}}\right\Vert_2^2 }& 0\\0&\frac{1}{\left\Vert\widetilde{\mathbf{h}}\right\Vert_2^2 }
		\end{bmatrix} \begin{bmatrix}
			\Re\left\{\beta^*\left(\frac{\partial \widetilde{\mathbf{h}}}{\partial \theta}\right)^H   \widetilde{\mathbf{h}} 
			\right\}&\Re\left\{\beta^*\left(\frac{\partial \widetilde{\mathbf{h}}}{\partial r}\right)^H   \widetilde{\mathbf{h}} 
			\right\} \\ 
			-\Im\left\{\beta^*\left(\frac{\partial \widetilde{\mathbf{h}}}{\partial \theta}\right)^H   \widetilde{\mathbf{h}}\right\} & -\Im\left\{\beta^*\left(\frac{\partial \widetilde{\mathbf{h}}}{\partial r}\right)^H   \widetilde{\mathbf{h}}
			\right\} 
		\end{bmatrix}\\
		&=\vert\beta\vert^2\begin{bmatrix}
			\left\Vert\frac{\partial \widetilde{\mathbf{h}}}{\partial \theta}\right\Vert_2^2 - 	\frac{\left\vert\left(\frac{\partial \widetilde{\mathbf{h}}}{\partial \theta}\right)^H   \widetilde{\mathbf{h}} \right\vert^2}{\left\Vert\widetilde{\mathbf{h}}\right\Vert^2_2} &\Re\left\{\left(\frac{\partial \widetilde{\mathbf{h}}}{\partial \theta}\right)^H\frac{\partial \widetilde{\mathbf{h}}}{\partial r}
			\right\} -\frac{\Re\left\{\widetilde{\mathbf{h}}^H  \left(\frac{\partial \widetilde{\mathbf{h}}}{\partial \theta}\right)  \left(\frac{\partial \widetilde{\mathbf{h}}}{\partial r}\right)^H   \widetilde{\mathbf{h}}   \right\}}{\left\Vert\widetilde{\mathbf{h}}\right\Vert^2_2}\\
			\Re\left\{\left(\frac{\partial \widetilde{\mathbf{h}}}{\partial \theta}\right)^H\frac{\partial \widetilde{\mathbf{h}}}{\partial r}
			\right\} -\frac{\Re\left\{\widetilde{\mathbf{h}}^H  \left(\frac{\partial \widetilde{\mathbf{h}}}{\partial \theta}\right)  \left(\frac{\partial \widetilde{\mathbf{h}}}{\partial r}\right)^H   \widetilde{\mathbf{h}}   \right\}}{\left\Vert\widetilde{\mathbf{h}}\right\Vert^2_2} &\left\Vert\frac{\partial \widetilde{\mathbf{h}}}{\partial r}\right\Vert_2^2 - 	\frac{\left\vert\left(\frac{\partial \widetilde{\mathbf{h}}}{\partial r}\right)^H   \widetilde{\mathbf{h}} \right\vert^2}{\left\Vert\widetilde{\mathbf{h}}\right\Vert^2_2}
		\end{bmatrix}\\
		&\triangleq\vert\beta\vert^2 \overline{\mathbf{Q}},
	\end{aligned}
\end{equation} 
\end{figure*}
where $\overline{\mathbf{Q}}$ is defined as the normalized Fisher matrix which is irrelated to the signal strength but the array manifold.

Then, the CRBs w.r.t. $\theta$ and $r$ are given by
\begin{equation}\label{CRB_t}
	\begin{aligned}
		\textbf{CRB}_\theta=&\frac{\sigma_n^2}{2\vert\beta\vert^2}\frac{\left\Vert\widetilde{\mathbf{h}}\right\Vert^2_2\left\Vert\frac{\partial \widetilde{\mathbf{h}}}{\partial r}\right\Vert_2^2 - 	 {\left\vert\left(\frac{\partial \widetilde{\mathbf{h}}}{\partial r}\right)^H   \widetilde{\mathbf{h}} \right\vert^2} }{\left\Vert\widetilde{\mathbf{h}}\right\Vert^2_2 {\rm det}( {\overline{\mathbf{Q}}} )}, 
	\end{aligned}
\end{equation}
\begin{equation}\label{CRB_r}
	\begin{aligned}
		\textbf{CRB}_r=&\frac{\sigma_n^2}{2\vert\beta\vert^2}\frac{\left\Vert\widetilde{\mathbf{h}}\right\Vert^2_2\left\Vert\frac{\partial \widetilde{\mathbf{h}}}{\partial \theta}\right\Vert_2^2 - 	 {\left\vert\left(\frac{\partial \widetilde{\mathbf{h}}}{\partial \theta}\right)^H   \widetilde{\mathbf{h}} \right\vert^2} }{\left\Vert\widetilde{\mathbf{h}}\right\Vert^2_2 {\rm det}({\overline{\mathbf{Q}}})}. 
	\end{aligned}
\end{equation}

In the later sections, the closed-form CRB will be derived by considering 
two scenarios for $\widetilde{\mathbf{h}}$: 1) WSMS with the SW, and 2) WSMS with the HSPW assumption.

\section{CRBs for SW-WSMS}\label{CRB_SW}
\subsection{Array Layout and Manifolds}
Contiuning from Section \ref{Signal}, the specific array configuration with the SW-WSMS is described.
 $K$ subarrays, with each connected $M$ antennas, are deployed at the TX. The total number of antennas at the TX, denoted as $N_t$ and defined below Eqn. (\ref{Y}), equals $KM$. The intra-subarray spacing and inter-subarray spacing are represented as $d$ and $D\triangleq(M-1)d+D_0$, respectively. 
Furthermore, we can see that
the $m$-th element of the $k$-th subarray is located on $(0,0,(m-1)d+(k-1)D)$, where $m\in\{1,\cdots,M\}$, and $k\in\{1,\cdots,K\}$. Denoted by $\mathbf{g}_t\triangleq\mathbf{b}_t(r,\theta)$, the $(k(M-1)+m)$-th element of which follows  
\begin{equation}\label{bnt}
[\mathbf{b}_t(r,\theta)]_{k(M-1)+m}=\sqrt{\frac{1}{N_t}}e^{-j\frac{2\pi}{\lambda}\sqrt{r^2-2n_tr\sin\theta+n_t^2}},
\end{equation}
 where $\lambda$ is the antenna wavelength, $n_t\triangleq \frac{2k-K+1}{2}D+\frac{2m-M+1}{2}d$, $m\in\{0,\cdots,M-1\}$, and $k\in\{0,\cdots,K-1\}$.
 
The RX adopts a UA with an inter-element spacing of $d$ to align with the fully-digital beamforming architecture. The RX array manifold can be mathematically expressed as:
\begin{equation}\label{gr}
\mathbf{g}_r\triangleq \mathbf{a}_r(\phi)= \sqrt{\frac{1}{N_r}}\left[e^{j\frac{\pi}{\lambda}(-N_r+1)d\sin\phi},\cdots ,e^{j\frac{\pi}{\lambda}(N_r-1)d\sin \phi}\right]^T.
\end{equation} 
Recall Eqn. (\ref{phiR}), $[\mathbf{a}_r(\phi)]_{n_r}$ is further re-defined w.r.t. $\{\theta, r\}$:
\begin{equation}
[\mathbf{a}_r(r,\theta)]_{n_r}\triangleq\sqrt{\frac{1}{N_r}} e^{j\frac{\pi}{\lambda}\left(\frac{(2n_r-N_r+1)d r\sin \theta}{\sqrt{R^2+r^2-2 R r \cos \theta}}\right)}.
\end{equation}  
 
 \subsection{CRB Derivation}
The array manifolds in $\widetilde{\mathbf{h}}$ have been defined, and they can be substituted into Eqs. (\ref{CRB_t}) and (\ref{CRB_r}) to further derive the CRB solution. It is important to note that the derivation of the CRB involves performing calculations for the AMFs, specifically regarding the derivative of $\widetilde{\mathbf{h}}$.

\subsubsection{Array Manifold Functions}\label{AMF1}
Through this paper, we adopt the orthogonal training matrix such that\footnote{In practical systems, $T\le N_t$ or sparse arrays can be considered to reduce the training time.} $\widetilde{\mathbf{F}}^*\widetilde{\mathbf{F}}^T=\mathbf{I}_{N_t}$. Besides, it is easy to know the 2-norm of the array manifold equals to $1$, i.e., $\left\Vert \mathbf{a}_r(r,\theta) \right\Vert^2_2=\left\Vert \mathbf{b}_t(r,\theta) \right\Vert^2_2=1$. As a result, the calculations regarding the derivative of $\widetilde{\mathbf{h}}$ are described as Eqs. (\ref{C1})-(\ref{C2}).
\begin{equation}\label{C1}
	\begin{aligned}
		&	\left\Vert \frac{\partial \widetilde{\mathbf{h}}}{\partial \theta} 	\right\Vert^2_2= \left\Vert \frac{ \mathbf{b}_t^*(r,\theta)}{\partial \theta} \right\Vert^2_2 \left\Vert \mathbf{a}_r(r,\theta) \right\Vert^2_2+ \left\Vert \frac{\partial \mathbf{a}_r(r,\theta)}{\partial \theta} \right\Vert^2_2 \left\Vert  \mathbf{b}_t^*(r,\theta) \right\Vert^2_2\\& \ \ \  \ \ +2\Re\left\{  \frac{\partial \mathbf{b}_t^T(r,\theta) }{\partial \theta} \mathbf{b}_t^*(r,\theta) \mathbf{a}^H_r(r,\theta)\frac{\partial \mathbf{a}_r(r,\theta)}{\partial \theta} 
		\right\}  \\& = \left\Vert \frac{ \mathbf{b}_t^*(r,\theta)}{\partial \theta} \right\Vert^2_2 + \left\Vert \frac{\partial \mathbf{a}_r(r,\theta)}{\partial \theta} \right\Vert^2_2\\ &  \ \ \  \ \ +2\Re\left\{  \frac{\partial \mathbf{b}_t^T(r,\theta) }{\partial \theta} \mathbf{b}_t^*(r,\theta) \mathbf{a}^H_r(r,\theta)\frac{\partial \mathbf{a}_r(r,\theta)}{\partial \theta} 
		\right\}  . 
	\end{aligned} 
\end{equation}
Similarly, other AMFs are calculated as follows.
\begin{equation}
	\begin{aligned}
	&	\left\Vert \frac{\partial \widetilde{\mathbf{h}}}{\partial r} \right\Vert^2_2= \left\Vert \frac{ \mathbf{b}_t^*(r,\theta)}{\partial r} \right\Vert^2_2  + \left\Vert \frac{\partial \mathbf{a}_r(r,\theta)}{\partial r} \right\Vert^2_2\\ & \ \ \ \ \ +2\Re\left\{  \frac{\partial \mathbf{b}_t^T(r,\theta) }{\partial r} \mathbf{b}_t^*(r,\theta) \mathbf{a}^H_r(r,\theta)\frac{\partial \mathbf{a}_r(r,\theta)}{\partial r} 
		\right\} ,
	\end{aligned}
\end{equation}
\begin{equation}
	\begin{aligned}
	\left(\frac{\partial \widetilde{\mathbf{h}}}{\partial \theta}\right)^H   \widetilde{\mathbf{h}}&=\frac{\partial \mathbf{b}_t^T(r,\theta)}{\partial \theta} \mathbf{b}_t^*(r,\theta) + \frac{\partial \mathbf{a}^H_r(r,\theta)}{\partial \theta}\mathbf{a}_r(r,\theta),
	\end{aligned}
\end{equation}
\begin{equation}
\left(\frac{\partial \widetilde{\mathbf{h}}}{\partial r}\right)^H   \widetilde{\mathbf{h}}= \frac{\partial \mathbf{b}_t^T(r,\theta)}{\partial r} \mathbf{b}_t^*(r,\theta) + \frac{\partial \mathbf{a}^H_r(r,\theta)}{\partial r}\mathbf{a}_r(r,\theta),
\end{equation}
\begin{equation}\label{C2}
	\begin{aligned}
		\left(\frac{\partial \widetilde{\mathbf{h}}}{\partial \theta}\right)^H\frac{\partial \widetilde{\mathbf{h}}}{\partial r}=&  \frac{\partial \mathbf{b}_t^T(r,\theta)}{\partial \theta} \frac{\partial \mathbf{b}_t^*(r,\theta)}{\partial r}+ \frac{\partial \mathbf{a}^H_r(r,\theta)}{\partial \theta}\frac{\partial \mathbf{a}_r(r,\theta)}{\partial r} \\
		&+ \frac{\partial \mathbf{b}_t^T(r,\theta)}{\partial \theta}\mathbf{b}_t^*(r,\theta) \mathbf{a}^H_r(r,\theta)\frac{\partial \mathbf{a}_r(r,\theta)}{\partial r}\\ &+ \mathbf{b}_t^T(r,\theta) \frac{\partial \mathbf{b}_t^*(r,\theta)}{\partial \theta} \frac{\partial \mathbf{a}_r^H(r,\theta)}{\partial r} \mathbf{a}_r(r,\theta).
	\end{aligned}
\end{equation}

\subsubsection{Calculation of sum formulas} To further simplify Eqs. (\ref{C1})-(\ref{C2}),
 we first give the derivative expression of $\mathbf{b}_t(r,\theta)$ w.r.t. $\theta$ and $r$ as
\begin{equation}
	\left[\frac{\partial \mathbf{b}_t(r,\theta)}{\partial \theta}\right]_{k(M-1)+m}=j\frac{2\pi}{\lambda}\sqrt{\frac{1}{N_t}}e^{-j\frac{2\pi}{\lambda}\sqrt{r_{n_t}}}\frac{n_tr\cos\theta}{\sqrt{r_{n_t}}},
\end{equation}
\begin{equation}
\left[\frac{\partial \mathbf{b}_t(r,\theta)}{\partial r}\right]_{k(M-1)+m}=j\frac{2\pi}{\lambda}\sqrt{\frac{1}{N_t}}e^{-j\frac{2\pi}{\lambda}\sqrt{r_{n_t}}}\frac{n_t\sin\theta-r}{\sqrt{r_{n_t}}},
\end{equation}
where $r_{n_t}\triangleq{r^2-2n_tr\sin\theta+n_t^2}$, and $n_t$ is defined below Eqn. (\ref{bnt}).

For the RX array manifold, we have
\begin{equation}\label{R1}
	\left[\frac{\partial \mathbf{a}_r(r,\theta)}{\partial \theta}\right]_{n_r}=j\frac{\pi(2n_r-N_r+1)d}{\lambda\sqrt{N_r}} e^{j\frac{\pi}{\lambda}(2n_r-N_r+1)d\sin \phi}\frac{\partial \sin\phi}{\partial \theta}.
\end{equation}
Particularly, Eqn. (\ref{phiR}) provides the derivative of $\sin \phi$ w.r.t. $\theta$ and $r$ as follows.
\begin{equation}
	\frac{\partial \sin\phi(r,\theta)}{\partial \theta}=\frac{r \cos \theta\left(R^2+r^2-2R r \cos \theta\right)-R r^2\sin^2\theta}{\left(R^2+r^2-2 R r \cos \theta\right)^{3 / 2}},
\end{equation}
\begin{equation}\label{R2}
	\frac{\partial \sin\phi(r,\theta)}{\partial r}=	\frac{R \sin \theta(R-r \cos \theta)}{\left(R^2+r^2-2 R r \cos \theta\right)^{3 / 2}}.
\end{equation}

Based on the above, the sum formulas that help derive AMFs are derived as follows
\begin{equation}\label{SF1}
	\begin{aligned}
 &\left\Vert \frac{\partial \mathbf{b}_t^*(r,\theta)}{\partial \theta} \right\Vert^2_2=\frac{4\pi^2r^2\cos^2\theta }{N_t\lambda^2} \sum_{k=0}^{K-1}\sum_{m=0}^{M-1}\frac{n_t^2}{{r^2-2n_tr\sin\theta+n_t^2}} \\
&=\frac{4\pi^2r^2\cos^2\theta }{N_t\lambda^2} \\ & \times \sum_{k=-\frac{K-1}{2}}^{\frac{K-1}{2}}\sum_{m=\frac{M-1}{2}}^{\frac{M-1}{2}}  \frac{(kD+md)^2}{{r^2-2(kD+md)r\sin\theta+(kD+md)^2}} \\
&= \frac{4\pi^2r^2\cos^2\theta }{N_t\lambda^2}  \mathcal{S}_{\theta^2}.
	\end{aligned}
\end{equation}
Similarly, other sum formulas are given by
$
	\frac{\partial \mathbf{b}_t^T(r,\theta) }{\partial \theta} \mathbf{b}_t^*(r,\theta)= j\frac{2\pi r\cos\theta}{\lambda N_t} \mathcal{S}_{\theta}
$,
$
		\frac{\partial \mathbf{b}_t^T(r,\theta) }{\partial r} \mathbf{b}_t^*(r,\theta) = j\frac{2\pi}{\lambda N_t} \mathcal{S}_{r}$,
$
		\left\Vert \frac{\partial \mathbf{b}_t^*(r,\theta)}{\partial r} \right\Vert^2_2
	=\frac{4\pi^2}{N_t\lambda^2} \mathcal{S}_{r^2}
$, and
$
		\frac{\partial \mathbf{b}_t^T(r,\theta) }{\partial \theta}\frac{\partial \mathbf{b}_t^*(r,\theta) }{\partial r}=\frac{4\pi^2 r\cos\theta}{\lambda^2 N_t}\mathcal{S}_{\theta r}
$.

To derive the closed-form solutions for sum formulas $\left\{ {\mathcal{S}}_{\theta^2}, {\mathcal{S}}_{\theta}, {\mathcal{S}}_{r^2}, {\mathcal{S}}_{r}, {\mathcal{S}}_{\theta r}\right\}$, the Riemann sum is adopted for accurate approximation, which is described in the following proposition.

\begin{Proposition}\label{P_S}
By applying the midpoint Riemann sum, the analytical solutions for  $\left\{ {\mathcal{S}}_{\theta^2}, {\mathcal{S}}_{\theta}, {\mathcal{S}}_{r^2}, {\mathcal{S}}_{r}, {\mathcal{S}}_{\theta r}\right\}$ can be expressed in terms of the functions $\left\{G_{\theta^2}(x),G_\theta(x),G_{r}(x),G_{\theta r}(x)\right\}$ and the bounds $\{x_1,x_2,x_3,x_4\}$ for the integral transformed through the midpoint Riemann sum.  
The specific expressions for the functions $\left\{G_{\theta^2}(x),G_\theta(x),G_{r}(x),G_{\theta r}(x)\right\}$ can be found in Appendix \ref{XL}. Here, the bounds are defined as follows:
$x_1\triangleq-\frac{K}{2}\Delta_{D}-\frac{M}{2} \Delta_{d}$, $x_2\triangleq-\frac{K}{2}\Delta_{D}+\frac{M}{2} \Delta_{d}$, $x_3\triangleq\frac{K}{2}\Delta_{D}-\frac{M}{2} \Delta_{d}$, and $x_4\triangleq\frac{K}{2}\Delta_{D}+\frac{M}{2} \Delta_{d}$. In addition, the variables $\Delta_{d}\triangleq\frac{d}{r}$ and $\Delta_{D}\triangleq\frac{D}{r}$ are defined to simplify the expressions.
\end{Proposition}

\begin{Proof}
Please see Appendix \ref{XL}.
\end{Proof}


 \begin{figure*}
 	\centering 
 	\includegraphics[height=6.5cm,width=11cm]{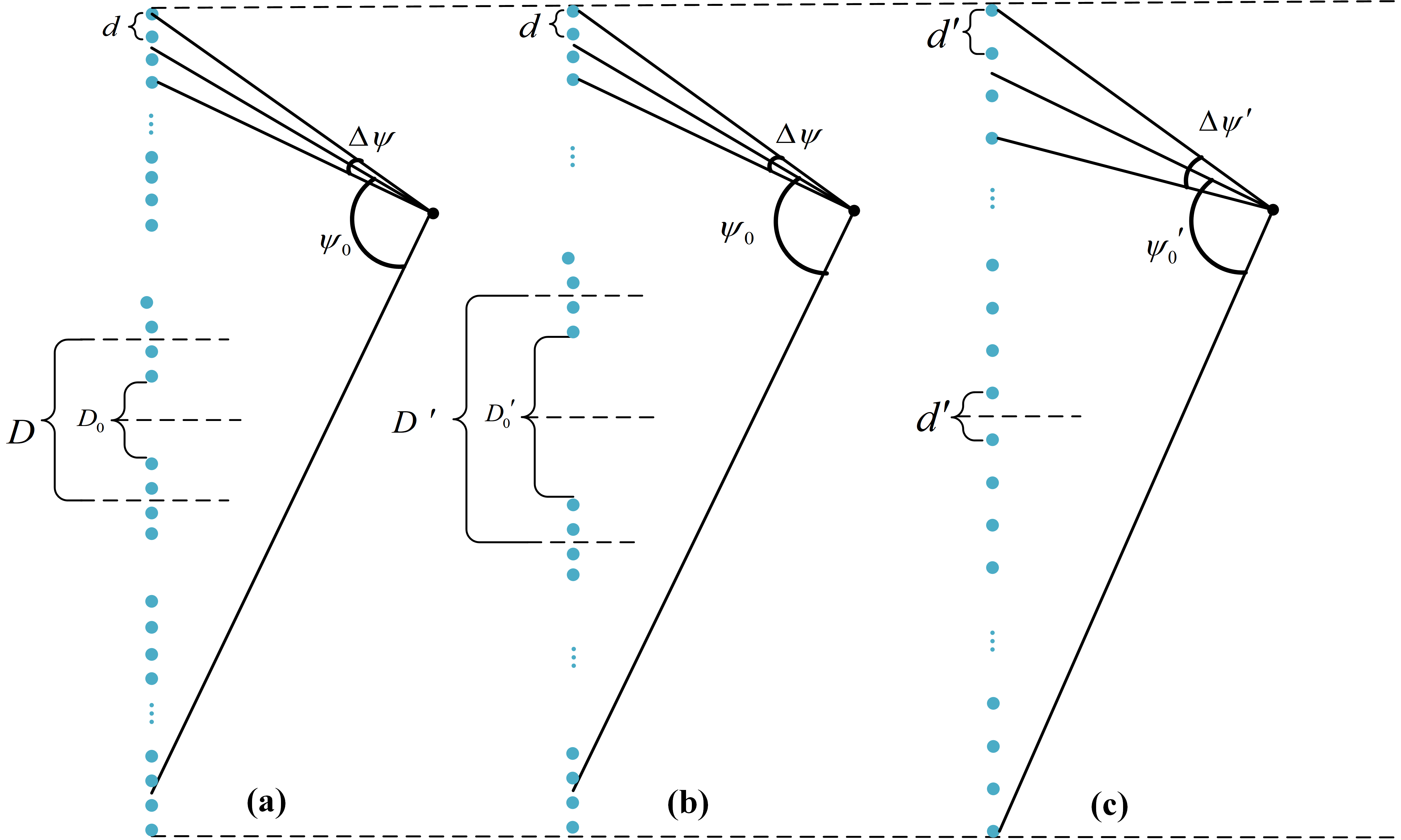}
 	\caption{The three different array layouts}\label{LAY} 
 \end{figure*} 
 
 The derivation of the closed-form solution holds significant importance in seeking the physical meaning behind the results. Consequently, the following proposition provides additional insights into the obtained solutions.
\begin{Proposition}\label{psi1}
	Let $\psi_0$ and $\Delta_\psi$ represent the angular spans of the WSMS, as shown in Fig. \ref{LAY}(a). With this notation, the sum formulas $\left\{ {\mathcal{S}}_{\theta^2}, {\mathcal{S}}_{\theta}, {\mathcal{S}}_{r^2}, {\mathcal{S}}_{r}, {\mathcal{S}}_{\theta r}\right\}$ can be re-written as functions of $\{\psi_0,\Delta_\psi\}$.
\end{Proposition}

\begin{Proof}
	\emph{It has been known that the functions $\left\{G_{\theta^2}(x),G_\theta(x),G_{r}(x),G_{\theta r}(x)\right\}$ in \textbf{Proposition \ref{P_S}} correspond to $\left\{ {\mathcal{S}}_{\theta^2}, {\mathcal{S}}_{\theta}, {\mathcal{S}}_{r^2}, {\mathcal{S}}_{r}, {\mathcal{S}}_{\theta r}\right\}$, where $x\in\{x_1,x_2,x_3,x_4\}$. Note also that these solution functions share certain common terms, such as ${1-2x\sin\theta+x^2}$ and $x$. 
As shown in Fig. \ref{TX-RX}, the cosine rule, given by $2xr^2\cos(\pi/2-\theta)=r^2+x^2-(r^\prime)^2$, and the sine rule, expressed as $\frac{{r}^\prime}{r}=\frac{\sin(\pi/2-\theta)}{\sin(\pi-(\pi/2-\theta)-\psi)}=\frac{\cos\theta}{\cos(\theta-\psi)}$, are applied. Hence 
\begin{equation}\label{12xi}
	1-2x\sin\theta+x^2=\frac{\cos^2\theta}{\cos^2(\theta-\psi)},
\end{equation} 
where
 $x$ can actually  be written as
\begin{equation}\label{xi}
x=\frac{\sin\psi}{\cos(\theta-\psi)}.
\end{equation} 
Noting that $\psi\in\{-\frac{\psi_0+\Delta_\psi}{2},-\frac{\psi_0-\Delta_\psi}{2},\frac{\psi_0-\Delta_\psi}{2},\frac{\psi_0+\Delta_\psi}{2}\}$ when $x\in\{x_1,x_2,x_3,x_4\}$.}
\end{Proof}
\begin{remark}
	\emph{Proposition \ref{psi1} highlights the significance of the physical angle, specifically the angular span, in determining the values of the AMFs as well as the CRB. This understanding allows us to investigate the AMF/CRB for array layouts with similar characteristics, from the perspective of the physical angle. Examples of such exploration include Corollary \ref{tar} and \textbf{Proposition \ref{UA}}.} 
\end{remark}

Based on the finding in \textbf{Proposition \ref{psi1}}, two useful corollaries are developed as follows. 
\begin{Corollary}\label{t0}
 \emph{When $\theta=0$, two important conclusions can be drawn: 1) $G_{\theta}(\psi,\theta=0)$ and $G_{\theta r}(\psi,\theta=0)$  are odd functions of $\psi$, hence it can be inferred that $\mathcal{S}_{\theta}=0$ and $\mathcal{S}_{\theta r}=0$ at $\theta=0$, and 2)  $G_{\theta^2}(\psi,\theta=0)$ and $G_{ r}(\psi,\theta=0)$  are even functions of $\psi$, hence the expressions for $\mathcal{S}_{\theta^2}$ and $\mathcal{S}_{r}$ can be simplified as follows: $\mathcal{S}_{\theta^2}=\frac{2}{\Delta_{D}\Delta_d}\left(G_{\theta^2}(\frac{\psi_0+\Delta_\psi}{2})-G_{\theta^2}(\frac{\psi_0-\Delta_\psi}{2})\right)$ and $\mathcal{S}_{r}=\frac{2}{\Delta_{D}\Delta_d}\left(G_{r}(\frac{\psi_0+\Delta_\psi}{2})-G_{r}(\frac{\psi_0-\Delta_\psi}{2})\right)$.}
\end{Corollary}

\begin{Corollary}\label{tar}
	\emph{Consider two WSMSs with the same angular span $\psi_0$ but different inter-subarray spacing and number of subarrays. Let us assume that the first WSMS has $K$ subarrays with an inter-subarray spacing of $D$, and the second WSMS has $K^\prime$ subarrays with an inter-subarray spacing of $D^\prime$. Importantly, the condition $KD=K^\prime D^\prime$ holds. 
		In this scenario, an interesting observation can be made: the ratio of their sum formulas is equal to the ratio of the number of subarrays, i.e., $\frac{\mathcal{S}}{\mathcal{S}^\prime}=\frac{K}{K^\prime}$,
		where $\mathcal{S}\in\left\{ {\mathcal{S}}_{\theta^2}, {\mathcal{S}}_{\theta}, {\mathcal{S}}_{r^2}, {\mathcal{S}}_{r}, {\mathcal{S}}_{\theta r}\right\}$ and $\mathcal{S}^\prime\in\left\{ {\mathcal{S}}_{\theta^2}^\prime, {\mathcal{S}}_{\theta}^\prime, {\mathcal{S}}_{r^2}^\prime, {\mathcal{S}}_{r}^\prime, {\mathcal{S}}_{\theta r}^\prime\right\}$ are the AMFs of the two WSMSs, respectively.
	}
\end{Corollary}

\begin{Proof}
	\emph{Let $\mathbf{b}^\prime_t(r,\theta)$ be the array manifold of the array, where
		the $(k(M-1)+m)$-th element of $\mathbf{b}^\prime_t(r,\theta)$ can be expressed by
		\begin{equation}\label{bnt2}
			[\mathbf{b}^\prime_t(r,\theta)]_{k^\prime(M-1)+m}=\sqrt{\frac{1}{K^\prime M}}e^{-j\frac{2\pi}{\lambda}\sqrt{r^2-2n^\prime_tr\sin\theta+(n^\prime_t)^2}},
		\end{equation}
		where $n_t^\prime\triangleq \frac{2k^\prime-K^\prime+1}{2}D+\frac{2m-M+1}{2}d$ and $m\in\{0,\cdots,M-1\}$,  $k^\prime\in\{0,\cdots,K^\prime-1\}$, $K^\prime=2$. 
		Recalling Eqn. (\ref{SF1}) and replacing $\mathbf{b}_t(r,\theta)$ with $\mathbf{b}^\prime_t(r,\theta)$, we can obtain 
		\begin{equation}
			\begin{aligned}
				\mathcal{S}^\prime_{\theta^2}&=\frac{1}{\Delta_d\Delta_{D^\prime}}	\sum_{k=-\frac{K^\prime-1}{2}}^{\frac{K^\prime-1}{2}} \left( F_{\theta^2}\left(k\Delta_{D^\prime}+\frac{M}{2} \Delta_{d}\right)\right. \\ & \ \ \ \ \ \ \ \ \ \ \ \ \ \ \ \ \ \ \ \  \left.
				-F_{\theta^2}\left(k\Delta_{D^\prime}-\frac{M}{2} \Delta_{d}\right)\right)\Delta_{D^\prime} \\
				&\approx
				\frac{1}{\Delta_d\Delta_{D^\prime}}\int_{\frac{K^\prime}{2}\Delta_{D^\prime}-\frac{M}{2} \Delta_{d}}^{\frac{K^\prime}{2}\Delta_{D^\prime}+\frac{M}{2} \Delta_{d}} F_{\theta^2}(x) {\rm d}x\\ &\ \ \ \ \ \  -\frac{1}{\Delta_d\Delta_{D^\prime}}\int_{-\frac{K^\prime}{2}\Delta_{D^\prime}-\frac{M}{2} \Delta_{d}}^{-\frac{K^\prime}{2}\Delta_{D^\prime}+\frac{M}{2} \Delta_{d}} F_{\theta^2}(x) {\rm d}x\\ &=
				\frac{1}{\Delta_d\Delta_{D^\prime}}\int_{\frac{K}{2}\Delta_{D}-\frac{M}{2} \Delta_{d}}^{\frac{K}{2}\Delta_{D}+\frac{M}{2} \Delta_{d}} F_{\theta^2}(x) {\rm d}x \\ & \ \ \ \ \ \ \ \  -
				\frac{1}{\Delta_d\Delta_D}\int_{-\frac{K}{2}\Delta_{D}-\frac{M}{2} \Delta_{d}}^{-\frac{K}{2}\Delta_{D}+\frac{M}{2} \Delta_{d}} F_{\theta^2}(x) {\rm d}x
				\\
				&=\frac{\Delta_{D}}{\Delta_{D^\prime}}\mathcal{S}_{\theta^2},
			\end{aligned}
		\end{equation}
		where $\Delta_{D^\prime}\triangleq \frac{D^\prime}{r}$.
		With $D^\prime=\frac{KD}{K^\prime}$ substituted, yielding $\frac{\mathcal{S}_{\theta^2}}{\mathcal{S}^\prime_{\theta^2}}=\frac{K}{K^\prime}$. This ratio can also be proved for other sum formulas.} 
\end{Proof}

Based on Eqs. (\ref{R1})-(\ref{R2}), the AMFs of the RX are provided as
$
		\left\Vert \frac{\partial \mathbf{a}_r(r,\theta)}{\partial \theta} \right\Vert^2_2=\left(\frac{\partial \sin\phi}{\partial \theta}\right)^2\frac{\pi^2d^2(N_r^2-1)}{3\lambda^2}
$,
$
		 \mathbf{a}^H_r(r,\theta) \frac{\partial \mathbf{a}_r(r,\theta)}{\partial \theta}  =0
$,
$
	\frac{\partial \mathbf{a}^H_r(r,\theta)}{\partial \theta} \frac{\partial \mathbf{a}_r(r,\theta)}{\partial r}= \frac{\pi^2 d^2(N_r^2-1)}{3\lambda^2  } \frac{\partial \sin\phi}{\partial \theta}\frac{\partial \sin\phi}{\partial r}
$.

\subsubsection{Closed-Form CRB}

Based on the derived AMFs, the entries of  $\overline{\mathbf{Q}}$ can be calculated:
\begin{equation}\label{h1}
	\begin{aligned}
  	&	\left\Vert\frac{\partial \widetilde{\mathbf{h}}}{\partial \theta}\right\Vert_2^2 - 	\frac{\left\vert\left(\frac{\partial \widetilde{\mathbf{h}}}{\partial \theta}\right)^H   \widetilde{\mathbf{h}} \right\vert^2}{\left\Vert\widetilde{\mathbf{h}}\right\Vert^2_2}= \left\Vert \frac{ \mathbf{b}_t^*(r,\theta)}{\partial \theta} \right\Vert^2_2     + \left\Vert \frac{\partial \mathbf{a}_r(r,\theta)}{\partial \theta} \right\Vert^2_2 \\& +2\Re\left\{  \frac{\partial \mathbf{b}_t^T(r,\theta) }{\partial \theta} \mathbf{b}_t^*(r,\theta) \mathbf{a}^H_r(r,\theta)\frac{\partial \mathbf{a}_r(r,\theta)}{\partial \theta} 
  		\right\} \\ &
  	-\left\vert \frac{\partial \mathbf{b}_t^T(r,\theta)}{\partial \theta} \mathbf{b}_t^*(r,\theta) + \frac{\partial \mathbf{a}^H_r(r,\theta)}{\partial \theta}\mathbf{a}_r(r,\theta)\right\vert^2 \\
  	=&\frac{4\pi^2r^2\cos^2\theta }{N_t\lambda^2} \left( \mathcal{S}_{\theta^2}  -  \mathcal{S}^2_{\theta}\right)+ \left(\frac{\partial \sin\phi}{\partial \theta}\right)^2\frac{\pi^2d^2(N_r^2-1)}{3\lambda^2}.
	\end{aligned}
\end{equation}
Similar this derivation, other entries can be given by 
$
[\overline{\mathbf{Q}}]_{2,2}=	\frac{4\pi^2 }{N_t\lambda^2}\left({ \mathcal{S}_{r^2}}-\frac{\mathcal{S}_r^2}{N_t}\right)+ \left(\frac{\partial \sin\phi}{\partial r}\right)^2\frac{\pi^2 d^2(N_r^2-1)}{3\lambda^2}
$, and
$
[\overline{\mathbf{Q}}]_{1,2} =\frac{4\pi^2r\cos\theta }{N_t\lambda^2} \left(\mathcal{S}_{\theta r}-\frac{\mathcal{S}_\theta\mathcal{S}_r}{N_t}\right)     + \left(\frac{\partial \sin\phi}{\partial \theta}\right) \left(\frac{\partial \sin\phi}{\partial r}\right)\frac{\pi^2 d^2(N_r^2-1)}{3\lambda^2}
$.

Consider the above, the following proposition gives the closed-form CRBs w.r.t. $\{\theta,r\}$.
\begin{Proposition}\label{CCRB0}
Denoted by $\chi_{N_r}\triangleq\frac{\pi^2 d^2(N_r^2-1)}{3\lambda^2}$, $\chi_{N_t}\triangleq\frac{4\pi^2r^2\cos^2\theta }{\lambda^2}$, $\phi_\theta\triangleq\frac{\partial \sin\phi}{\partial \theta}$, and $\phi_r\triangleq\frac{\partial \sin\phi}{\partial r}$, the closed-form CRBs are given by
\begin{equation}\label{CRBT}
	\begin{aligned}
		\textbf{CRB}_\theta=&\frac{\sigma_n^2}{2\vert\beta\vert^2}\frac{\frac{\chi_{N_t}}{r^2\cos^2\theta}\left(\frac{\mathcal{S}_{r^2}}{N_t}-\frac{\mathcal{S}_r^2}{N_t^2}\right)+ \chi_{N_r} \phi_r^2}{ {\rm det}({\overline{\mathbf{Q}}})},
	\end{aligned}
\end{equation}
\begin{equation}\label{CRBR}
	\begin{aligned}
		\textbf{CRB}_r=&\frac{\sigma_n^2}{2\vert\beta\vert^2}\frac{		\chi_{N_t} \left(\frac{\mathcal{S}_{\theta^2}}{N_t}-\frac{\mathcal{S}_\theta^2}{N_t^2}\right)+ \chi_{N_r} \phi_\theta^2}{ {\rm det}({\overline{\mathbf{Q}}})}, 
	\end{aligned}
\end{equation}
where the normalized Fisher matrix $	\overline{\mathbf{Q}}$ follows 
$[\overline{\mathbf{Q}}]_{1,1}=\chi_{N_t} \left(\frac{\mathcal{S}_{\theta^2}}{N_t}-\frac{\mathcal{S}_\theta^2}{N_t^2}\right)+ \chi_{N_r} \phi_\theta^2$, $[\overline{\mathbf{Q}}]_{1,2}=[\overline{\mathbf{Q}}]_{2,1}=
\frac{\chi_{N_t}}{r\cos\theta}\left(\frac{\mathcal{S}_{\theta r}}{N_t}-\frac{\mathcal{S}_\theta\mathcal{S}_r}{N_t^2}\right)+ \chi_{N_r}\phi_\theta\phi_r$,  $[\overline{\mathbf{Q}}]_{2,2}=\frac{\chi_{N_t}}{r^2\cos^2\theta}\left(\frac{\mathcal{S}_{r^2}}{N_t}-\frac{\mathcal{S}_r^2}{N_t^2}\right)+ \chi_{N_r} \phi_r^2$,
and $\left\{ {\mathcal{S}}_{\theta^2}, {\mathcal{S}}_{\theta}, {\mathcal{S}}_{r^2}, {\mathcal{S}}_{r}, {\mathcal{S}}_{\theta r}\right\}$ can be found in \textbf{Proposition \ref{P_S}}.
\end{Proposition}


\begin{Corollary}\label{KKP}
\emph{Recalling Corollary \ref{tar}, which establishes a linear ratio between the sum formulas of the two array layouts depicted in Fig. \ref{LAY}(a) and (b), we observe that they have the same normalized Fisher matrix. This implies that the difference in their CRBs is solely reflected in the received SNR. Consequently, the ratio $\frac{K}{K^\prime}$ mentioned in Corollary \ref{tar} also applies to the CRBs of the two array layouts. In other words, we have $\frac{\textbf{CRB}^\prime_{\theta/r}}{\textbf{CRB}_{\theta/r}}=\frac{K}{K^\prime}$, where $\textbf{CRB}^\prime_{\theta/r}$ represents the angle/range CRB of layout 2. 
}
\end{Corollary}

\begin{Proof}
	\emph{Denoted by $\overline{\mathbf{Q}}^\prime$ the CRB matrix of layout 2 in Fig. \ref{LAY}(b), similar to the derivation of $\overline{\mathbf{Q}}$, we have $[\overline{\mathbf{Q}}^\prime]_{1,1}=	\chi_{N_t} \left(\frac{\mathcal{S}^\prime_{\theta^2}}{N_t^\prime}-\frac{(\mathcal{S}_\theta^\prime)^2}{(N_t^\prime)^2}\right)+ \chi_{N_r} \phi_\theta^2$, $[\overline{\mathbf{Q}}^\prime]_{1,2}=[\overline{\mathbf{Q}}^\prime]_{2,1}=
		\frac{\chi_{N_t}}{r\cos\theta}\left(\frac{\mathcal{S}^\prime_{\theta r}}{N_t^\prime}-\frac{\mathcal{S}^\prime_\theta\mathcal{S}^\prime_r}{(N_t^\prime)^2}\right)+ \chi_{N_r}\phi_\theta\phi_r $, and $[\overline{\mathbf{Q}}^\prime]_{2,2}=\frac{\chi_{N^\prime_t}}{r^2\cos^2\theta}\left(\frac{\mathcal{S}^\prime_{r^2}}{N^\prime_t}-\frac{(\mathcal{S}_r^\prime)^2}{(N_t^\prime)^2}\right)+ \chi_{N_r} \phi_r^2$,
where $N_t^\prime\triangleq K^\prime M$. Then, according to the linear ratio in \emph{Corollary \ref{tar}},
 the equation holds as
 \begin{equation}
\frac{\mathcal{S}^\prime_{\theta^2}}{N_t^\prime}=\frac{\frac{K^\prime}{K}\mathcal{S}_{\theta^2}}{K^\prime M}=\frac{\mathcal{S}_{\theta^2}}{N_t}.
 \end{equation} 
Similarly, $\left\{\frac{\mathcal{S}^\prime_{\theta}}{N_t^\prime},\frac{\mathcal{S}^\prime_{r}}{N_t^\prime},\frac{\mathcal{S}^\prime_{\theta r}}{N_t^\prime}, \frac{\mathcal{S}^\prime_{r^2}}{N_t^\prime}\right\}$ are equal to $\left\{\frac{\mathcal{S}_{\theta}}{N_t},\frac{\mathcal{S}_{r}}{N_t},\frac{\mathcal{S}_{\theta r}}{N_t},\frac{\mathcal{S}_{r^2}}{N_t}\right\}$, respectively.
Moreover, noting that $\beta$ is different for the two array layouts, where $\beta=\alpha \sqrt{N_rN_t}$ for layout 1 and $\beta^\prime=\alpha \sqrt{N_rN_t^\prime}$ for layout 2. Hence 
\begin{equation}
\frac{\textbf{CRB}_{\theta}^\prime}{\textbf{CRB}_{\theta}}=\frac{\textbf{CRB}_{r}^\prime}{\textbf{CRB}_{r}}=\frac{|\beta|^2}{|\beta^\prime|^2}=\frac{K}{K^\prime}.
\end{equation}
It can be observed that increasing the number of subarrays in layout 2 leads to a linear improvement in CRB performance. Additionally, we can deduce that when $D^\prime < \frac{KD}{K^\prime}$, the ratio $\frac{\textbf{CRB}_{\theta/r}^\prime}{\textbf{CRB}_{\theta/r}} > \frac{K}{K^\prime}$. This implies that increasing the inter-subarray spacing can decrease the CRB, resulting in improved performance. }
\end{Proof}

\begin{Corollary}\label{CRBTR}
\emph{When $\theta=0$, according to \textbf{Proposition \ref{psi1}} and Corollary \ref{t0}, the closed-form CRBs can be represented by $\psi_0$ and $\Delta_\psi$:
	\begin{equation}\label{CRBT0}
		\begin{aligned}
		&	\textbf{CRB}_\theta(\psi_0,\Delta_\psi,\theta=0)=\\& \ \ \ \ \ \ \frac{\sigma_n^2}{2\vert\beta\vert^2}\frac{1}{\frac{4\pi^2r^2}{\lambda^2}\frac{\mathcal{S}_{\theta^2}(\psi_0,\Delta_\psi,\theta=0)}{N_t}+ \frac{\pi^2 d^2(N_r^2-1)}{3\lambda^2} \frac{r^2}{(R-r)^2}}, 
		\end{aligned}
	\end{equation}
	\begin{equation}\label{CRBR0}
		\begin{aligned}
			&\textbf{CRB}_r(\psi_0,\Delta_\psi,\theta=0)=\\ & \ \ \ \ \ \ \frac{\sigma_n^2}{2\vert\beta\vert^2}\frac{1}{\frac{4\pi^2}{\lambda^2}\left(\frac{\mathcal{S}_{r^2}(\psi_0,\Delta_\psi,\theta=0)}{N_t}-\frac{(\mathcal{S}_r(\psi_0,\Delta_\psi,\theta=0))^2}{N_t^2}\right)},
		\end{aligned}
	\end{equation}
	where $\mathcal{S}_{\theta^2}(\psi_0,\Delta_\psi,\theta=0)$, $\mathcal{S}_{r^2}(\psi_0,\Delta_\psi,\theta=0)$, and $\mathcal{S}_r(\psi_0,\Delta_\psi,\theta=0)$ can be obtained by Corollary \ref{t0}.  
} 
\end{Corollary}

\begin{Proof}
\emph{Recalling Corollary \ref{t0}, we have
discussed the case of $\theta=0$ for calculating $\left\{ {\mathcal{S}}_{\theta^2}, {\mathcal{S}}_{\theta}, {\mathcal{S}}_{r^2}, {\mathcal{S}}_{r}, {\mathcal{S}}_{\theta r}\right\}$. From this analysis, important results are obtained:  $\mathcal{S}_{\theta}=\mathcal{S}_{\theta r}=0$,    $\mathcal{S}_{\theta^2}=\frac{2}{\Delta_{D}\Delta_d}\left(G_{\theta^2}(\frac{\psi_0+\Delta_\psi}{2})-G_{\theta^2}(\frac{\psi_0-\Delta_\psi}{2})\right)$, and $\mathcal{S}_{r}=\frac{2}{\Delta_{D}\Delta_d}\left(G_{r}(\frac{\psi_0+\Delta_\psi}{2})-G_{r}(\frac{\psi_0-\Delta_\psi}{2})\right)$.
Additionally, according to Eqn. (\ref{R2}), $\phi_r=0$ at $\theta=0$. 
Thus, considering $\overline{\mathbf{Q}}$ at $\theta=0$ can derive
 Eqs. (\ref{CRBT0}) and (\ref{CRBR0}).}
\end{Proof}
 \begin{remark}
\emph{According to \emph{Corollary \ref{CRBTR}}, some obvious conclusions regarding the RX parameter can be observed. It is worth noting that in Eqn. (\ref{CRBT0}), as $r$ approaches $R$, the $\textbf{CRB}_\theta$ tends to 0. Moreover, when $\theta=0$, it follows that $\phi_r=0$, resulting in the normalized range CRB being independent of the RX parameter.}
 \end{remark}

 We are also interested in comparing the WSMS with the UA layout, which has the same aperture and number of antennas. Directly analyzing the difference in CRBs between these two array layouts is challenging. Therefore, we leverage the subarray structure to compare the CRBs of WSMSs and UAs. First, it is important to note that when the inter-subarray spacing $D_0$ is equal to the intra-subarray spacing $d$, the WSMS degenerates into a UA.
 \begin{Proposition}\label{UA}
 	We consider the WSMS and UA layouts with the same array aperture and antenna number. For the UA, the inter-element spacing is set to $d^\prime=\frac{D(K-1)+d(M-1)}{KM-1}$, as shown in Fig. \ref{LAY}(c). We find that when $\theta=0$, the CRBs of the two layouts satisfy  $\textbf{CRB}^{\rm WSMS}_\theta<	\textbf{CRB}^{\rm UA}_\theta$\footnote{Here, we focus on the special case of $\theta=0$ for the angle CRB, and the theoretical analysis for general cases of angle/range CRB remains open for future exploration.}.
 \end{Proposition}

\begin{Proof}
\emph{Despite the uniform distribution of antenna elements in UAs, we consider the $K$ subarray structure with $M$ elements in each subarray and the inter-subarray spacing $Md^\prime$, where $d^\prime\triangleq\frac{D(K-1)+d(M-1)}{KM-1}$. According to \textbf{Proposition \ref{psi1}} and Corollary \ref{t0}, we can use the angular spans $\psi_0^\prime$ and $\Delta_{\psi}^\prime$ to characterize the CRB. As shown in Fig. \ref{LAY}, it can be observed that $\psi_0^\prime+\Delta_\psi^\prime=\psi_0+\Delta_\psi$ and $\psi_0^\prime-\Delta_\psi^\prime<\psi_0-\Delta_\psi$. Recalling Eqn. (\ref{CRBT0}), our aim is to prove $\mathcal{S}_{\theta^2}(\psi_0,\Delta_\psi,\theta=0)>\mathcal{S}_{\theta^2}(\psi_0^\prime,\Delta_\psi^\prime,\theta=0)$, which is equivalent to proving
	\begin{equation}\label{GE}
		\begin{aligned}
		&	G_{\theta^2}\left(\frac{\psi_0+\Delta_\psi}{2}\right)-	G_{\theta^2}\left(\frac{\psi_0-\Delta_\psi}{2}\right) \\ &  \ \ \ -	G_{\theta^2}\left(\frac{\psi_0^\prime+\Delta_\psi^\prime}{2}\right)+	G_{\theta^2}\left(\frac{\psi_0^\prime-\Delta_\psi^\prime}{2}\right)>0.
		\end{aligned}
	\end{equation}
Since when $\theta=0$, $\frac{{\rm d} G_{\theta^2}(\psi)}{{\rm d} \psi}=2\psi-\frac{1}{2}\tan\psi-\psi\sec^2\psi<0$ at $\psi\in(0,\frac{\pi}{2}]$, $G_{\theta^2}(\psi)$ is a monotonically decreasing function on this interval. Considering $\psi_0^\prime+\Delta_\psi^\prime=\psi_0+\Delta_\psi$ and $\psi_0^\prime-\Delta_\psi^\prime<\psi_0-\Delta_\psi$, we conclude that Eqn. (\ref{GE}) holds true. 
}
\end{Proof}
\section{CRBs for HSPW-WSMS}\label{CRB-HSPW}
Although the array manifold $\mathbf{b}_t(r,\theta)$ provides an accurate characterization of the near-field parameters, its complex expression renders it impractical for real-world systems. In this regard, the HSPW assumption is favored. Hence, the objective of this section is to derive closed-form expressions for the CRB under the HSPW assumption and compare them with the CRBs derived in the previous section.
\subsection{Array Manifolds}
Under the HSPW assumption, the intra-subarray and inter-subarray manifolds exhibit spherical and planar wavefronts, respectively.  
 Thus, the TX array manifold $\mathbf{g}_t$ can be constructed by $\mathbf{g}_t\triangleq\mathbf{w}(r,\theta)\otimes \mathbf{a}_t(\theta)$, with $\mathbf{a}_t(\theta)\in\mathbb{C}^{M\times 1}$ and $\mathbf{w}_t(\theta,r)\in\mathbb{C}^{K\times 1}$ denoting the far-field and near-field array responses, respectively. The expression for $\mathbf{a}_t(\theta)$ is given by
\begin{equation}
	\begin{aligned}
		\mathbf{a}_t(\theta)\triangleq&\sqrt{\frac{1}{M}}[e^{-j\frac{\pi}{\lambda}(M-1)d\sin\theta},\cdots,e^{j\frac{\pi}{\lambda}(2m-M+1)d\sin \theta}, \\  &\ \ \ \ \ \  \ \ \ \  \ \cdots ,e^{j\frac{\pi}{\lambda}(M-1)d\sin \theta}]^T.
	\end{aligned}
\end{equation}
In addition, the $k$-th element of $\mathbf{w}(r,\theta)$ is expressed by
\begin{equation}\label{bnk}
	[\mathbf{w}(r,\theta)]_k=\sqrt{\frac{1}{K}}e^{-j\frac{2\pi}{\lambda}\sqrt{r^2-2n_kr\sin\theta+n_k^2}},
\end{equation}
where $n_k\triangleq \frac{2k-K+1}{2}D$, $k\in\{0,\cdots,K-1\}$.

\subsection{CRB Derivation}

\subsubsection{Array Manifold Functions}
Substituting the array manifold into Eqn. (\ref{wh}), we obtain the expression for $\widetilde{\mathbf{h}}$ in the case of HSPW-based WSMS as follows:
\begin{equation}
\widetilde{\mathbf{h}}\triangleq\left(\widetilde{\mathbf{F}}^T\left(\mathbf{w}^*(r,\theta)\otimes\mathbf{a}^*_t(\theta)\right)\right)\otimes\mathbf{a}_r(r,\theta).
\end{equation}

As Section \ref{AMF1}, the orthogonal training matrix meets  $\widetilde{\mathbf{F}}^*\widetilde{\mathbf{F}}^T=\mathbf{I}_{N_t}$. Besides, we can know that the 2-norm of the array manifold equals to $1$, i.e., $\left\Vert \mathbf{a}_r(r,\theta) \right\Vert^2_2=\left\Vert \mathbf{a}_t(\theta) \right\Vert^2_2=\left\Vert \mathbf{w}(r,\theta) \right\Vert^2_2=1$. Therefore, the calculations regarding the derivative of $\widetilde{\mathbf{h}}$ are described as Eqs. (\ref{WHT1})-(\ref{WHT2}):
\begin{equation}\label{WHT1}
	\begin{aligned}
		\left\Vert \frac{\partial \widetilde{\mathbf{h}}}{\partial \theta} 	\right\Vert^2_2&= \left\Vert \frac{ \mathbf{w}^*(r,\theta)}{\partial \theta} \right\Vert^2_2+\left\Vert \frac{ \mathbf{a}_t^*(\theta)}{\partial \theta} \right\Vert^2_2 + \left\Vert \frac{\partial \mathbf{a}_r(r,\theta)}{\partial \theta} \right\Vert^2_2 \\ & +2\Re\left\{  \frac{\partial \mathbf{w}^T(r,\theta) }{\partial \theta} \mathbf{w}^*(r,\theta) \mathbf{a}^H_r(r,\theta)\frac{\partial \mathbf{a}_r(r,\theta)}{\partial \theta} \right\}   \\
		&   +2\Re\left\{  \frac{\partial \mathbf{w}^T(r,\theta) }{\partial \theta} \mathbf{w}^*(r,\theta) \mathbf{a}^T_t(\theta)\frac{\partial \mathbf{a}^*_t(\theta)}{\partial \theta}
		\right\}\\&   +2\Re\left\{  \frac{\partial \mathbf{a}^T_t(\theta) }{\partial \theta} \mathbf{a}^*_t(\theta) \mathbf{a}^H_r(r,\theta)\frac{\partial \mathbf{a}_r(r,\theta)}{\partial \theta}
		\right\},  
	\end{aligned}
\end{equation}
\begin{equation}
	\begin{aligned}
		\left\Vert \frac{\partial \widetilde{\mathbf{h}}}{\partial r} 	\right\Vert^2_2&=2\Re\left\{  \frac{\partial \mathbf{w}^T(r,\theta) }{\partial r} \mathbf{w}^*(r,\theta) \mathbf{a}^H_r(r,\theta)\frac{\partial \mathbf{a}_r(r,\theta)}{\partial r} \right\}\\ &   \ \ \ + \left\Vert \frac{ \mathbf{w}^*(r,\theta)}{\partial r} \right\Vert^2_2 + \left\Vert \frac{\partial \mathbf{a}_r(r,\theta)}{\partial r} \right\Vert^2_2,  
	\end{aligned}
\end{equation}
\begin{equation}
	\begin{aligned}
	\left(\frac{\partial \widetilde{\mathbf{h}}}{\partial \theta}\right)^H   \widetilde{\mathbf{h}}&= \frac{\partial \mathbf{w}^T(r,\theta)}{\partial \theta} \mathbf{w}^*(r,\theta) +  \frac{\partial \mathbf{a}_t^T(\theta)}{\partial \theta} \mathbf{a}_t^*(\theta) \\ & \ \ \  + \frac{\partial \mathbf{a}^H_r(r,\theta)}{\partial \theta}\mathbf{a}_r(r,\theta),
	\end{aligned}
\end{equation}
\begin{equation}
	\begin{aligned}
		\left(\frac{\partial \widetilde{\mathbf{h}}}{\partial r}\right)^H   \widetilde{\mathbf{h}}&= \frac{\partial \mathbf{w}^T(r,\theta)}{\partial r} \mathbf{w}^*(r,\theta)  + \frac{\partial \mathbf{a}^H_r(r,\theta)}{\partial r}\mathbf{a}_r(r,\theta),
	\end{aligned}
\end{equation}
\begin{equation}\label{WHT2}
	\begin{aligned}
		&	\left(\frac{\partial \widetilde{\mathbf{h}}}{\partial \theta}\right)^H\frac{\partial \widetilde{\mathbf{h}}}{\partial r}= \frac{\partial \mathbf{w}^T(r,\theta)}{\partial \theta}\frac{\partial \mathbf{w}^*(r,\theta)}{\partial r} +\frac{\partial \mathbf{w}^T(r,\theta)}{\partial \theta}\mathbf{w}^*(r,\theta)\\ &\ \ \ \times \mathbf{a}^H_r(r,\theta) \frac{\partial \mathbf{a}_r(r,\theta)}{\partial r} + \mathbf{w}^T(r,\theta)\frac{\partial \mathbf{w}^*(r,\theta)}{\partial r} \frac{\partial \mathbf{a}_t^T(\theta)}{\partial \theta}\mathbf{a}_t^*(\theta)\\&\ \ \ + \frac{\partial \mathbf{a}_t^T(\theta)}{\partial \theta}\mathbf{a}_t^*(\theta) \mathbf{a}^H_r(r,\theta) \frac{\partial \mathbf{a}_r(r,\theta)}{\partial r}+\mathbf{w}^T(r,\theta)\frac{\partial \mathbf{w}^*(r,\theta)}{\partial r}\\&\ \ \ \times \frac{\partial \mathbf{a}^H_r(r,\theta)}{\partial \theta}\mathbf{a}_r(r,\theta)+\frac{\partial \mathbf{a}^H_r(r,\theta)}{\partial \theta}\frac{\partial \mathbf{a}_r(r,\theta)}{\partial r}.
	\end{aligned}
\end{equation}
\subsection{Calculation of sum formulas} 
To further derive Eqs. (\ref{WHT1})-(\ref{WHT2}),
we first give the derivative expression of $\mathbf{a}_t(\theta)$ and $\mathbf{w}(r,\theta)$ w.r.t. $\theta$ and $r$ as
\begin{equation}
	\left[\frac{\partial \mathbf{a}_t(\theta)}{\partial \theta}\right]_m=j\frac{\pi(2m-M+1)d\cos\theta}{\lambda}\left[\mathbf{a}_t(\theta)\right]_m,
\end{equation}
\begin{equation}
\left[\frac{\partial \mathbf{w}(r,\theta)}{\partial \theta}\right]_k=j\frac{2\pi}{\lambda}\frac{n_kr\cos\theta}{\sqrt{r_{k}}}\left[\mathbf{w}(r,\theta)\right]_k,
\end{equation}
\begin{equation}
	\left[\frac{\partial \mathbf{w}(r,\theta)}{\partial r}\right]_k=j\frac{2\pi}{\lambda}\frac{n_k\sin\theta-r}{\sqrt{r_{k}}}\left[\mathbf{w}(r,\theta)\right]_k,
\end{equation}
 where $m\in\{1,\cdots,M\}$, $k\in\{1,\cdots,K\}$, and
  $r_{k}\triangleq r^2-2n_kr\sin\theta+n_k^2$.

Since the TX array manifold $\mathbf{g}_t$ is the Kronecker product of $\mathbf{w}(r,\theta)$ and $a_t(\theta)$, its deriatives can be expressed by
\begin{equation}
\frac{\partial\mathbf{g}_t }{\partial \theta}=\mathbf{w}(r,\theta)\otimes \frac{\partial \mathbf{a}_t(\theta)}{\partial \theta}+\frac{\partial \mathbf{w}(r,\theta)}{\partial \theta}\otimes\mathbf{a}_t(\theta),
\end{equation}
 \begin{equation}
 	\frac{\partial\mathbf{g}_t }{\partial r}=\frac{\partial \mathbf{w}(r,\theta)}{\partial r}\otimes\mathbf{a}_t(\theta).
 \end{equation}

According to this, the TX AMFs w.r.t. $\frac{\partial\mathbf{g}_t }{\partial \theta}$ and $\frac{\partial\mathbf{g}_t }{\partial r}$ can be expressed by $\frac{\partial \mathbf{w}(r,\theta)}{\partial r}$, $\frac{\partial \mathbf{w}(r,\theta)}{\partial \theta}$, and $\frac{\partial \mathbf{a}_t(\theta)}{\partial \theta}$. In this case, the TX sum formulas can be obtained following the similar derivation of Eqn. (\ref{SF1}):
$
		\left\Vert \frac{ \mathbf{w}^*(r,\theta)}{\partial \theta} \right\Vert^2_2=\frac{4\pi^2r^2\cos^2\theta}{\lambda^2K} \widetilde{\mathcal{S}}_{\theta^2}
$,
$
		\left\Vert \frac{\partial \mathbf{w}^*(r,\theta)}{\partial r} \right\Vert^2_2
	= \frac{4\pi^2}{K\lambda^2}\widetilde{\mathcal{S}}_{r^2}
$,
$
\frac{\partial \mathbf{w}^T(r,\theta) }{\partial \theta} \mathbf{w}^*(r,\theta)=j\frac{2\pi r\cos\theta}{\lambda K} \widetilde{\mathcal{S}}_\theta
$,
$
	\frac{\partial \mathbf{w}^T(r,\theta) }{\partial r} \mathbf{w}^*(r,\theta)	= j\frac{2\pi}{\lambda K} \widetilde{\mathcal{S}}_{r}
$, and
$
		 \frac{\partial \mathbf{w}^T(r,\theta) }{\partial \theta}\frac{\partial \mathbf{w}^*(r,\theta) }{\partial r}=\frac{4\pi^2 r\cos\theta}{\lambda^2 K}\widetilde{\mathcal{S}}_{\theta r}
$.
 Furthermore, $\left\Vert \frac{\partial \mathbf{a}_t(\theta)}{\partial \theta} \right\Vert^2_2=\frac{\pi^2d^2\cos^2\theta(M^2-1)}{3\lambda^2}$ and $	\mathbf{a}^H_t(r,\theta) \frac{\partial \mathbf{a}_t(r,\theta)}{\partial \theta}=0$ can be easily obtained.
  
  Similar to \textbf{Proposition \ref{P_S}}, 	$\left\{\widetilde{\mathcal{S}}_{\theta^2},\widetilde{\mathcal{S}}_{\theta},\widetilde{\mathcal{S}}_{r^2},\widetilde{\mathcal{S}}_{r},\widetilde{\mathcal{S}}_{\theta r}\right\}$ can be accurately solved by the midpoint Riemann sum, as explained in Appendix \ref{WS}. 
  However, the results in Appendix \ref{WS} has a limit insight for the CRB parameters. However, the insights provided by the results in Appendix \ref{WS} are limited in terms of the CRB parameters. In fact, these parameters can be better understood and represented by the angular span $\psi_0$, as described in the following proposition.
 \begin{Proposition}\label{WS0}
 	Based on \textbf{\textbf{Proposition \ref{psi1}}} and \emph{Corollary \ref{t0}},  $\left\{\widetilde{\mathcal{S}}_{\theta^2},\widetilde{\mathcal{S}}_{\theta},\widetilde{\mathcal{S}}_{r^2},\widetilde{\mathcal{S}}_{r},\widetilde{\mathcal{S}}_{\theta r}\right\}$ can be represented by the span angular $\psi_0$. In particular, when $\theta=0$, the simplified expressions w.r.t. $\psi_0$ are given by
\begin{equation}\label{SP1}
	\begin{aligned}
		\widetilde{\mathcal{S}}_{\theta^2}(\psi_0,\theta=0)=K-\frac{K\psi_0}{2\tan\frac{\psi_0}{2}},
	\end{aligned}
\end{equation}
\begin{equation}\label{SP11}
	\begin{aligned}
		\widetilde{\mathcal{S}}_{r^2}(\psi_0,\theta=0)=\frac{K\psi_0}{2\tan\frac{\psi_0}{2}},
	\end{aligned}
\end{equation}
 \begin{equation}\label{SP2}
 	\begin{aligned}
 		\widetilde{\mathcal{S}}_{r}(\psi_0,\theta=0)
 		= -\frac{K}{2\tan\frac{\psi_0}{2}}	\ln\frac{1+\sin\frac{\psi_0}{2}}{1-\sin\frac{\psi_0}{2}},
 	\end{aligned}
 \end{equation}
 \begin{equation}
 	\begin{aligned}
 		\widetilde{\mathcal{S}}_{\theta r}(\psi_0,\theta=0)=		\widetilde{\mathcal{S}}_\theta(\psi_0,\theta=0)=0.
 	\end{aligned}
 \end{equation}
  \end{Proposition}

 \begin{Proof}
\emph{By invoking Eqn. (\ref{WST2}) for example, we can examine the case of $\theta=0$, resulting in the expression:
$
		\widetilde{\mathcal{S}}_{\theta^2}(\theta=0)=K-\frac{1}{\Delta_D}\tan^{-1}\left(\frac{K\Delta_D}{2}\right)+\frac{1}{\Delta_D}\tan^{-1}\left(-\frac{K\Delta_D}{2}\right).
$
Then, substituting $\frac{K\Delta_D}{2}=\psi_0$ to obtain Eqn. (\ref{SP1}). Similarly, Eqs. (\ref{SP11})-(\ref{SP2}) can be derived.
}
 \end{Proof}

 \begin{Corollary}\label{asym1}
 \emph{When $\psi_0\rightarrow  \pi $, indicating that $\frac{KD}{r}\rightarrow \infty$,
the asymptotic AMFs are given by: $	\widetilde{\mathcal{S}}_{\theta^2}(\psi_0\rightarrow \pi,\theta=0)=K$, $	\widetilde{\mathcal{S}}_{r^2}(\psi_0\rightarrow \pi,\theta=0)=0$, and $	\widetilde{\mathcal{S}}_{r}(\psi_0\rightarrow \pi,\theta=0)=0$.
}
 \end{Corollary}

 \begin{Corollary}\label{asym2}
  \emph{When $\psi_0\rightarrow  0$, indicating that $\frac{KD}{r}\rightarrow 0$, the asymptotic AMFs are given by: $	\widetilde{\mathcal{S}}_{\theta^2}(\psi_0\rightarrow 0,\theta=0)=0$, $	\widetilde{\mathcal{S}}_{r^2}(\psi_0\rightarrow 0,\theta=0)=K$, and $	\widetilde{\mathcal{S}}_{r}(\psi_0\rightarrow 0,\theta=0)=-K$.
 }
 \end{Corollary}

\subsection{Closed-Form CRBs}

Utilizing the previously derived AMFs and sum formulas, we can derive the expressions for each entry of the Fisher matrix $	\widetilde{\overline{\mathbf{Q}}}^\prime$, yielding  
$
\left[\widetilde{\overline{\mathbf{Q}}}^\prime\right]_{1,1}= \frac{4\pi^2r^2\cos^2\theta}{\lambda^2K}\left(\widetilde{\mathcal{S}}_{\theta^2}-\frac{\widetilde{\mathcal{S}}^2_\theta}{K}\right)+ \frac{\pi^2 d^2(M^2-1)}{3\lambda^2}\cos^2\theta+ \frac{\pi^2 d^2(N_r^2-1)}{3\lambda^2}\phi_\theta^2
$,
$
	\left[\widetilde{\overline{\mathbf{Q}}}^\prime\right]_{2,2}=\frac{4\pi^2}{\lambda^2K} \left(\widetilde{\mathcal{S}}_{r^2}-\frac{\widetilde{\mathcal{S}}^2_r}{K}\right)+\frac{\pi^2 d^2(N_r^2-1)}{3\lambda^2}\phi_r^2
$,
and $
\left[\widetilde{\overline{\mathbf{Q}}}^\prime\right]_{1,2}= \frac{4\pi^2 r\cos\theta}{\lambda^2 K}\left( \widetilde{\mathcal{S}}_{\theta r}-\frac{\widetilde{\mathcal{S}}_\theta  \widetilde{\mathcal{S}}_{r}}{K}\right)+\frac{\pi^2 d^2(N_r^2-1)}{3\lambda^2}\phi_\theta\phi_r$.
 Then, the closed-form CRBs w.r.t. $\theta$ and $r$, with the HPSW-based WSMS, are concluded.
\begin{Proposition}\label{CCRB}
Denoted by ${\chi}_{K}\triangleq\frac{4\pi^2r^2\cos^2\theta }{\lambda^2}$ and ${\chi}_{M}\triangleq\frac{\pi^2 d^2(M^2-1)}{3\lambda^2}$, the closed-form CRBs are given by 
\begin{equation}\label{CRBT2}
	\begin{aligned}
		\widetilde{\textbf{CRB}}_\theta=&\frac{\sigma_n^2}{2\vert\beta\vert^2}\frac{\frac{\chi_K}{r^2\cos^2\theta}\left(\frac{{ \widetilde{\mathcal{S}}_{r^2}}}{K}-\frac{\widetilde{\mathcal{S}}_r^2}{K^2}\right)+ \chi_{N_r} \phi_r^2}{ {\rm det}({\widetilde{\overline{\mathbf{Q}}}^\prime})}, 
	\end{aligned}
\end{equation}
\begin{equation}\label{CRBR2}
	\begin{aligned}
		\widetilde{\textbf{CRB}}_r=&\frac{\sigma_n^2}{2\vert\beta\vert^2}\frac{\chi_K \left(\frac{\widetilde{\mathcal{S}}_{\theta^2}}{K}-\frac{\widetilde{\mathcal{S}}_\theta^2}{K^2}\right)+\chi_M \cos^2\theta +\chi_{N_r}\phi_\theta^2}{ {\rm det}({\widetilde{\overline{\mathbf{Q}}}^\prime})},
	\end{aligned}
\end{equation}
where $\chi_{N_r}$, $\phi_\theta$, and $\phi_r$ are defined in \textbf{Proposition \ref{CCRB0}}, and $[\widetilde{\overline{\mathbf{Q}}}^\prime]_{1,1}=\chi_K \left(\frac{\widetilde{\mathcal{S}}_{\theta^2}}{K}-\frac{\widetilde{\mathcal{S}}_\theta^2}{K^2}\right)+\chi_M \cos^2\theta +\chi_{N_r}\phi_\theta^2$, $[\widetilde{\overline{\mathbf{Q}}}^\prime]_{1,2}=[\widetilde{\overline{\mathbf{Q}}}^\prime]_{2,1}=\frac{\chi_K}{r\cos\theta}\left(\frac{\widetilde{\mathcal{S}}_{\theta r}}{K}-\frac{\widetilde{\mathcal{S}}_\theta\widetilde{\mathcal{S}}_r}{K^2}\right)+ \chi_{N_r}\phi_\theta\phi_r $, and $[\widetilde{\overline{\mathbf{Q}}}^\prime]_{2,2}=\frac{\chi_K}{r^2\cos^2\theta}\left(\frac{{ \widetilde{\mathcal{S}}_{r^2}}}{K}-\frac{\widetilde{\mathcal{S}}_r^2}{K^2}\right)+ \chi_{N_r} \phi_r^2$.
\end{Proposition}

\begin{remark}\label{diff}
\emph{In comparison to the SW-based CRBs presented in \textbf{Proposition \ref{CCRB0}}, the HSPW-based CRBs can be decomposed into the SW and PW components. $\chi_{N_t}$ is transformed into $\{\chi_K,\chi_M\}$, and $\frac{\mathcal{S}_{\theta^2}}{N_t}-\frac{\mathcal{S}_{\theta}^2}{N_t^2}$ is transformed into $\left\{\frac{\widetilde{\mathcal{S}}{\theta^2}}{K}-\frac{\widetilde{\mathcal{S}}{\theta}^2}{K^2},\cos^2\theta\right\}$. It is important to note that the PW component exists only in $\left[\widetilde{\overline{\mathbf{Q}}}^\prime\right]_{1,1}$, and it significantly affects the angle CRB. This is because the PW array manifold factor does not depend on the range and therefore has a limited contribution to range sensing. 
}
\end{remark}
 
\begin{Corollary}
\emph{Similar to Corollary \ref{KKP}, the CRBs of the two layouts with HSPW also satisfy the linear ratio:
	\begin{equation}
		\frac{\widetilde{\textbf{CRB}}_{\theta}^\prime}{\widetilde{\textbf{CRB}}_{\theta}}=\frac{\widetilde{\textbf{CRB}}_{r}^\prime}{\widetilde{\textbf{CRB}}_{r}}=\frac{K}{K^\prime}.
	\end{equation}
}
\end{Corollary}

\begin{Corollary}\label{asym3}
\emph{According to \textbf{Proposition \ref{psi1}} and Corollary \ref{t0}, the closed-form CRB can be represented by $\psi_0$ at $\theta=0$. 
	\begin{equation}\label{WCRBT0}
		\begin{aligned}
			&\widetilde{\textbf{CRB}}_\theta(\psi_0,\theta=0)=\\ &   \frac{\sigma_n^2\lambda^2}{8\vert\beta\vert^2\pi^2}\frac{1}{r^2-\frac{r^2\psi_0}{2\tan\frac{\psi_0}{2}} + \frac{ d^2(M^2-1)}{12}+\frac{ d^2(N_r^2-1)}{12} \frac{r^2}{(R-r)^2}}, 
		\end{aligned}
	\end{equation}
	\begin{equation}\label{WCRBR0}
		\begin{aligned}
		&	\textbf{CRB}_r(\psi_0,\theta=0)=\\&\ \ \ \ \ \frac{\sigma_n^2\lambda^2}{8\vert\beta\vert^2\pi^2}\frac{1}{
				\left(\frac{\psi_0}{2\tan\frac{\psi_0}{2}}-\frac{1}{2\tan^2\frac{\psi_0}{2}}	\left(\ln\frac{1+\sin\frac{\psi_0}{2}}{1-\sin\frac{\psi_0}{2}}\right)^2\right)},
		\end{aligned}
	\end{equation}
	where $\mathcal{S}_{\theta^2}(\psi_0,\theta=0)$, $\mathcal{S}_{r^2}(\psi_0,\theta=0)$, and $\mathcal{S}_r(\psi_0,\theta=0)$ can be found in Eqs. (\ref{SP1})-(\ref{SP2}).
} 
\end{Corollary}

\begin{Proof}
\emph{Based on \textbf{Proposition \ref{WS0}}, and $\phi_r=0$ at $\theta=0$, we can derive $	\widetilde{\overline{\mathbf{Q}}}$ at $\theta=0$ to obtain
 Eqs. (\ref{WCRBT0}) and (\ref{WCRBR0}).}
\end{Proof}

In light of \emph{Corollary \ref{asym1}, \ref{asym2}, and \ref{asym3}}, the asymptotic CRBs at $\theta=0$ are derived as follows.  

\begin{Corollary} 
\emph{When $\psi_0\rightarrow \pi$, the asymptotic CRBs are given by
	\begin{equation}\label{WCRBT}
	\begin{aligned}
		&\widetilde{\textbf{CRB}}_\theta(\psi_0\rightarrow\pi,\theta=0)=\\ & \ \ \ \ \ \frac{\sigma_n^2\lambda^2}{8\vert\beta\vert^2\pi^2}\frac{1}{r^2+ \frac{ d^2(M^2-1)}{12}+\frac{ d^2(N_r^2-1)}{12} \frac{r^2}{(R-r)^2}},
	\end{aligned}
	\end{equation} 
	\begin{equation}
		\widetilde{\textbf{CRB}}_r(\psi_0\rightarrow\pi,\theta=0)\rightarrow \infty.
	\end{equation}}
\end{Corollary}

\begin{Corollary}
\emph{When $\psi_0\rightarrow 0$, the asymptotic CRBs are given by
	\begin{equation}\label{WCRBR}
		\widetilde{\textbf{CRB}}_\theta(\psi_0\rightarrow 0,\theta=0)=\frac{\sigma_n^2\lambda^2}{8\vert\beta\vert^2\pi^2}\frac{1}{ \frac{ d^2(M^2-1)}{12}+\frac{ d^2(N_r^2-1)}{12} \frac{r^2}{(R-r)^2}},
	\end{equation} 
	\begin{equation}
		\widetilde{\textbf{CRB}}_r(\psi_0\rightarrow 0,\theta=0)\rightarrow \infty.
\end{equation} }
\end{Corollary}


\section{Simulation Results}\label{SR}
In this section, we present the results of our simulations aimed at investigating the impact of various factors on angle/range CRB performance for WSMSs. The general parameter setting is described as follows: the system frequency is $100$ GHz, the transmit SNR $\frac{1}{\sigma^2_n}$ is set to $0$ dB, $M=128$, and $D_0\triangleq 2^I\cdot\frac{\lambda}{2}$ with $I$ controlling the inter-subarray spacing. Particularly, when $I=0$, the WSMS degenerates to the DUA.
\begin{itemize}
\item \textbf{SW-WSMS}: Directly calculating the sum formulas for the SW-WSMS.
\item \textbf{SW-WSMS Approx.}: Approximating the sum formulas using the midpoint Riemann sum for the SW-WSMS (closed-form Eqns. (\ref{CRBT}) and (\ref{CRBR})).  
\item \textbf{HSPW-WSMS}: Directly calculating the sum formulas for the HSPW-WSMS.
\item \textbf{HSPW-WSMS Approx.}: Approximating the sum formulas using the midpoint Riemann sum for the HSPW-WSMS (closed-form Eqns. (\ref{CRBT2}) and (\ref{CRBR2})).  
\item \textbf{PW-WSMS}: Directly calculating the sum formulas for the PW-WSMS.
\item \textbf{SW-UA}: Directly calculating the sum formulas for the SW-UA.
\item \textbf{SW-DUA}: Directly calculating the sum formulas for the SW-DUA.
\end{itemize}
 \begin{figure}
	\centering
	\subfigure[Root $\textbf{CRB}_{\theta}$]{
		\includegraphics[width=2.62in]{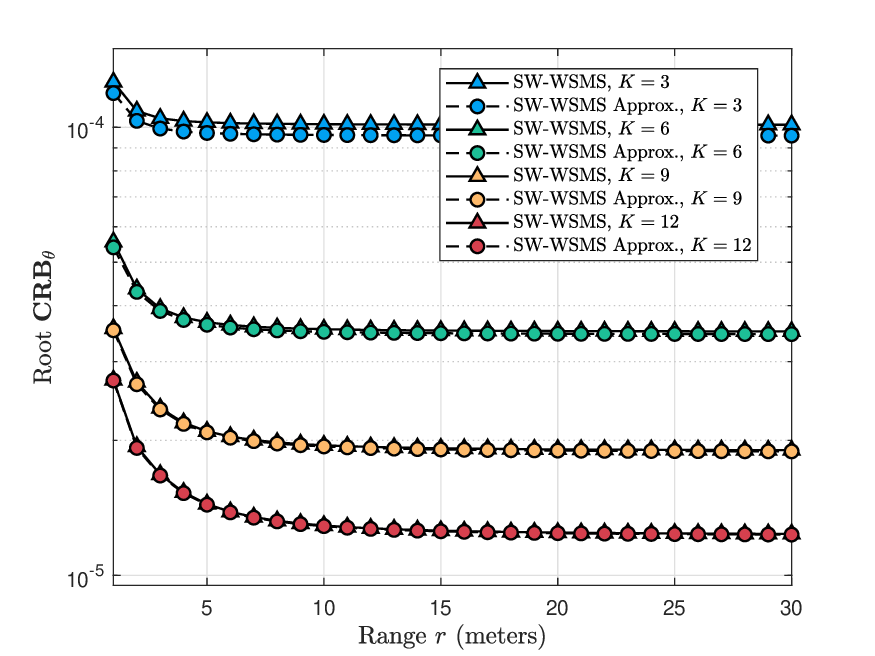}
	}
	\quad    
	\subfigure[Root $\textbf{CRB}_{r}$]{
		\includegraphics[width=2.62in]{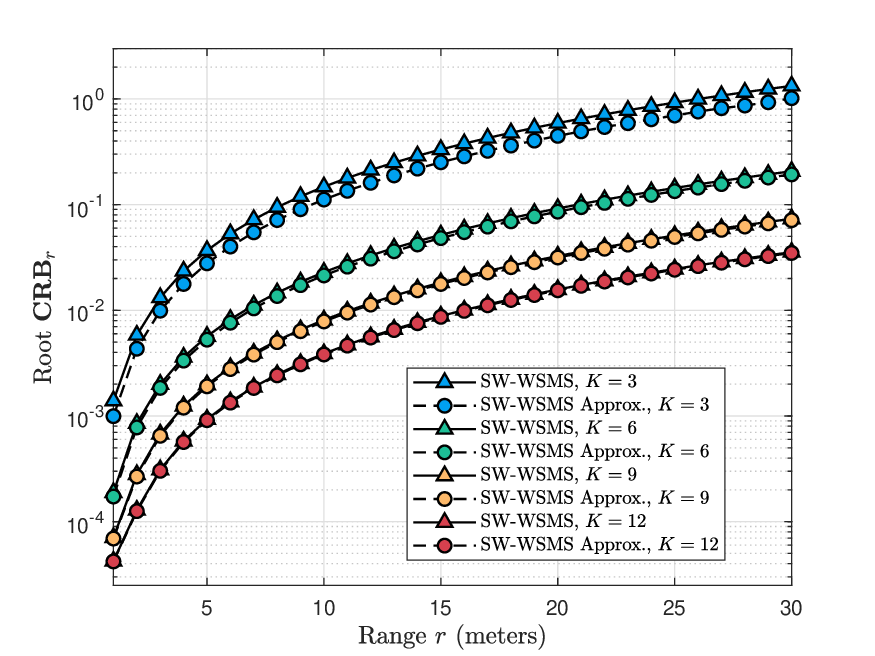}
	}
	\caption{The root CRBs of SW-WSMS and SW-WSMS Approx. with different $K$ and $r$.}
	\label{Approx}
\end{figure}

Given that the closed-form CRBs have some degree of approximation due to the approximate integral transformed by the Riemann sum, we initially focus on evaluating the approximation error. For this purpose, we set $N_r=1$ and $\theta=\pi/4$, with $K\in\{3,6,9,12\}$, where $N_r$ is set to $1$ to disregard the influence of RX on CRB.
In Fig. \ref{Approx} (a) and (b), presented results depict the examination of the range $r\in[2,50]$. It is observed that as $r$ increases, the range CRB also increases, while the angle CRB displays the opposite behavior by decreasing as $r$ increases. This disparity is attributed to the opposing impact of the near-field effect on angle and range estimation. As $r$ grows larger, the SW faces challenges in accurately sensing the range, resulting in an increase in the range CRB. Notably, the angle and range CRBs of SW-HSPW and SW-HSPW Approx. converge as $K$ increases. When $K$ is small, there exists a marginal error between the CRBs of these two schemes, which can be neglected.
Overall, SW-HSPW Approx. proves to be sufficiently accurate in approximating SW-HSPW, especially when $K$ is not small. This also applies for HSPW-HSPW Approx. and HSPW-HSPW.
\begin{figure}
	\centering
	\subfigure[Root $\textbf{CRB}_\theta$, $\theta=\pi/4$]{
		\includegraphics[width=2.62in]{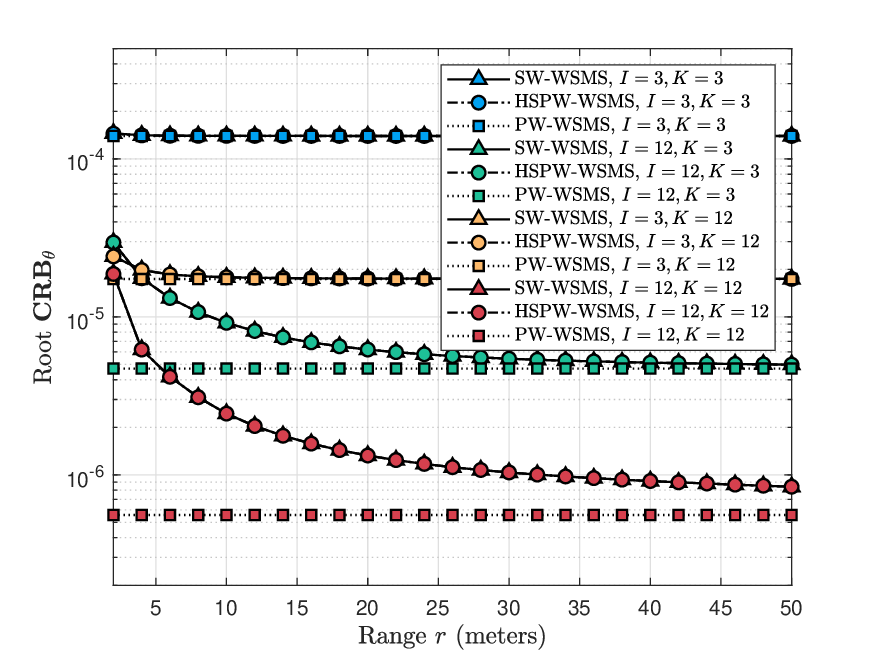}
	}
	\quad    
	\subfigure[Root $\textbf{CRB}_r$, $\theta=\pi/4$]{
		\includegraphics[width=2.62in]{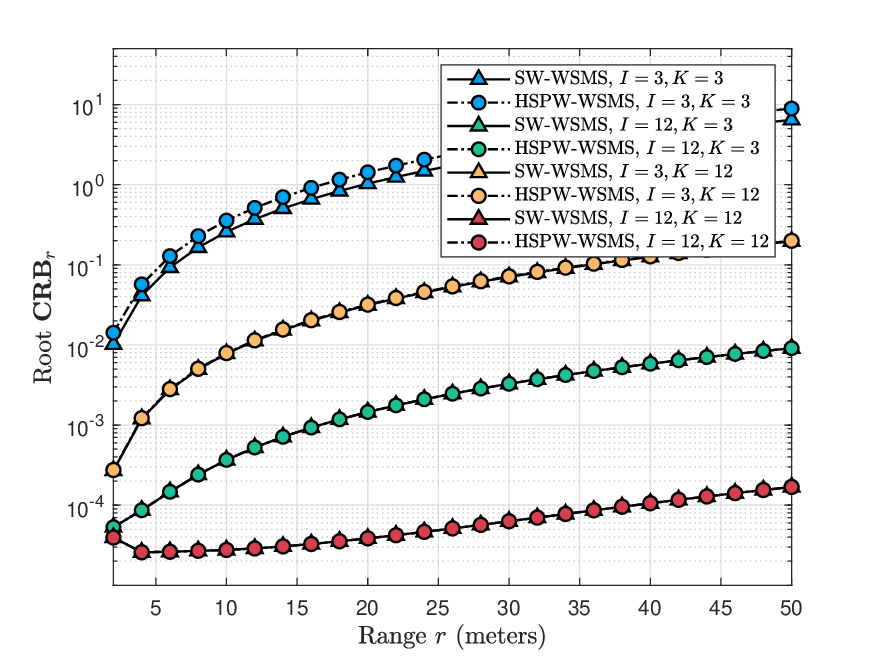}
	}
	\caption{The root CRBs of SW-/HSPW-/PW-WSMW with different $\{K,I,r\}$.}
	\label{RDK}
\end{figure}

\begin{figure}
 	\centering
 	\subfigure[Root $\textbf{CRB}_\theta$, $r=10$]{
 		\includegraphics[width=2.62in]{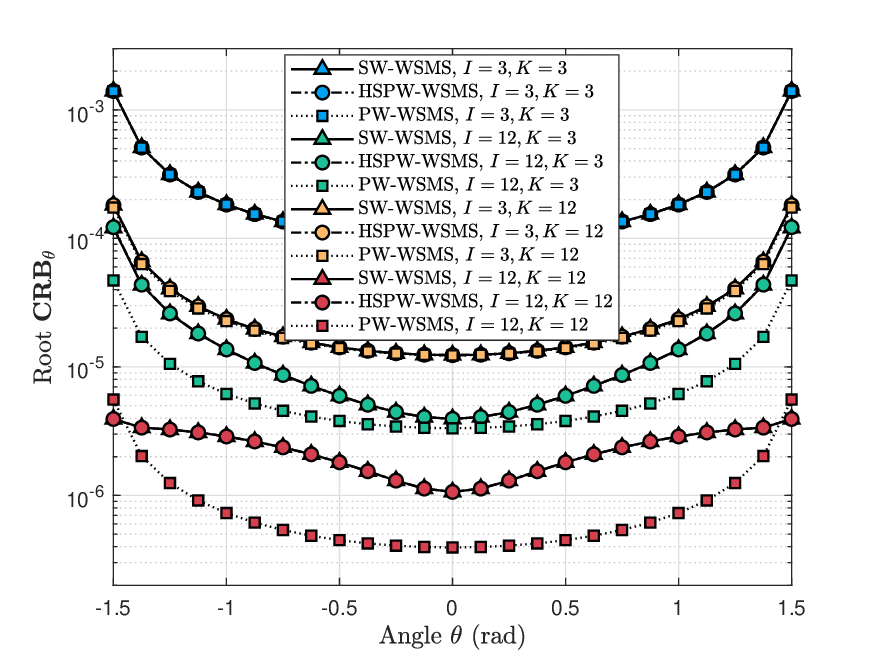}
 	}
 	\quad    
 	\subfigure[Root $\textbf{CRB}_r$, $r=10$]{
 		\includegraphics[width=2.62in]{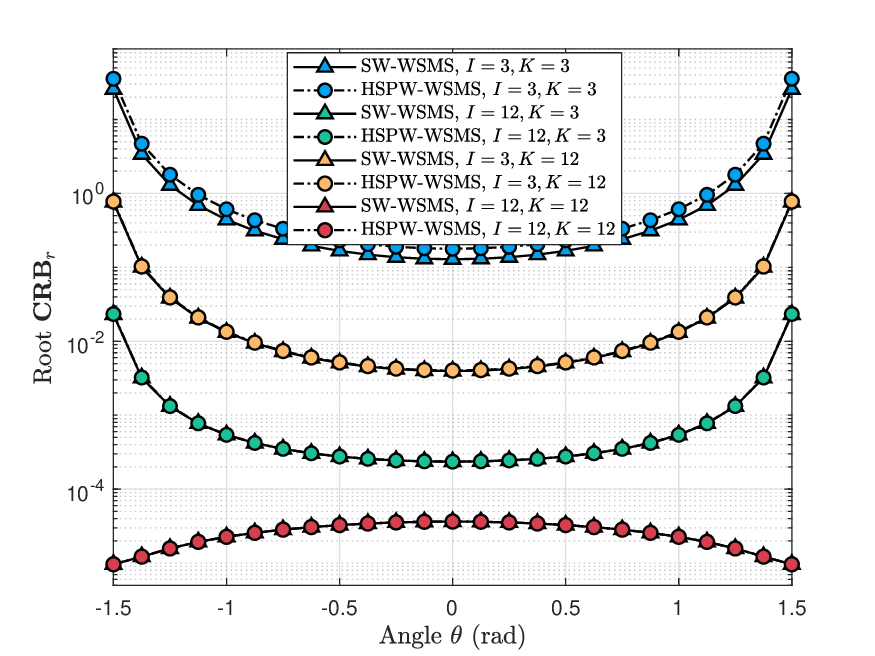}
 	}
 	\caption{The root CRBs of SW-/HSPW-/PW-WSMS with different $\{K,I,\theta\}$.}
 	\label{ADK}
 \end{figure}

Due to the inevitable error caused by the model assumption, specifically PW-WSMS and HSPW-WSMS, it is essential to evaluate the magnitude of this error by comparing their CRBs with the CRB of the SW-WSMS. In this scenario, the parameters are set as $N_r=1$, $\theta\in[-1.5,1.5]$, and $r\in[2,50]$. Additionally, four cases are considered regarding the parameters ${I,K}$: 1) $I=3$ and $K=3$, 2) $I=12$ and $K=3$, 3) $I=3$ and $K=12$, and 4) $I=12$ and $K=12$.
In Fig. \ref{RDK} (a), the root $\text{CRB}_\theta$ is evaluated with different parameters. When $\{K,I\}$ are small, such as the blue curves, the $\textbf{CRB}_\theta$ exhibits an unclear change with respect to the range $r$. Consequently, the three schemes, SW-WSMS, HSPW-WSMS, and PW-WSMS, yield consistent results in terms of $\textbf{CRB}_\theta$. However, as $\{K,I\}$ increase, a noticeable error arises between HSPW-/SW-WSMS and PW-WSMS. This error diminishes as the range $r$ increases, indicating that the PW-WSMS model exhibits significant error in the near-field region when the array aperture is large. Furthermore, it can be observed that the angle CRBs of the SW-WSMS and HSPW-WSMS maintain a consistent trend regardless of changes in the array aperture. A slightly different observation arises for $\textbf{CRB}_r$. In Fig. \ref{RDK} (b), when $\{K,I\}$ are small (the blue curve), a slight error can be observed between the SW-WSMS and HSPW-WSMS, which decreases as the array aperture increases. Conversely, the CRBs vary with $\theta$ in Fig. \ref{ADK} (a) and (b) yield similar conclusions. As depicted in Fig. \ref{ADK}, a significant error is evident between the  SW-/HSPW-WSMS and PW-WSMS when ${I,K}$ are large. Additionally, for small ${I,K}$, a slight error is present between the SW-WSMS and HSPW-WSMS in terms of root $\textbf{CRB}_r$, as shown in Fig. \ref{ADK} (b). Notably, when large ${I,K}$ are utilized (the red curve), an opposite trend is observed compared to the other curves. This phenomenon suggests that when the array aperture is sufficiently large, the range CRB exhibits a decreasing trend with respect to $|\theta|$.

\begin{figure}
	\centering 
	\includegraphics[height=4.7cm,width=6.5cm]{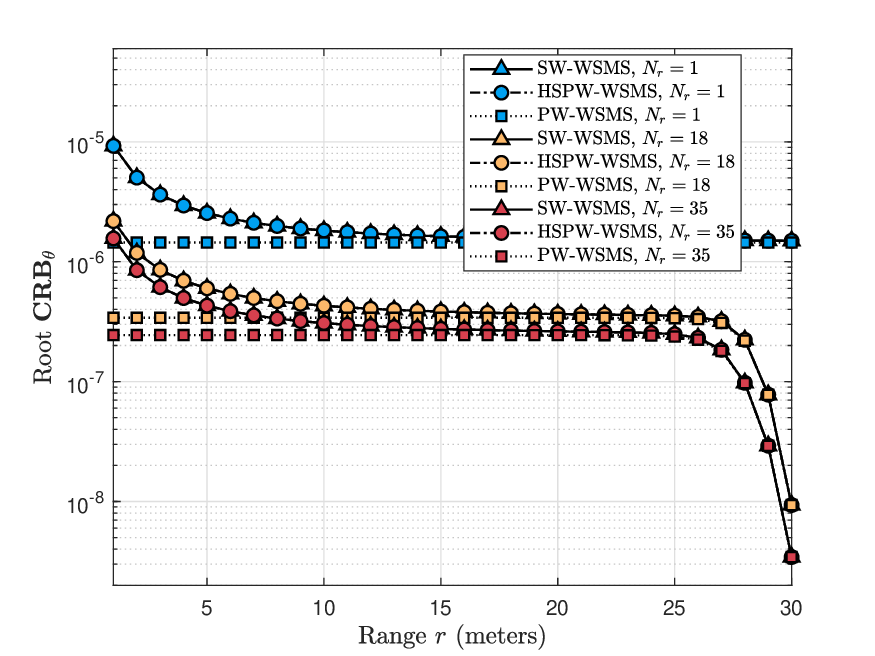}
	\caption{The angle CRB versus the range $r$.}\label{RXr}
\end{figure}

All the simulations mentioned above assume $N_r=1$ to neglect the impact of the RX end. However, we have observed that the RX parameters have a significant effect on the angle CRB. Therefore, here we aim to explore the influence of the RX parameter using the parameter set $\{N_r\in\{1,18,35\},\theta=0,r\in[1,30],R=31,K=12,I=10\}$, as depicted in Fig. \ref{RXr}. In particular, we note a distinct difference between the cases of $N_r=1$ and $N_r>1$ in terms of $\textbf{CRB}_\theta$ as $r$ approaches $R$. This disparity arises because when $N_r>1$ and $r$ approaches $R$, the RX parameter in Eqn. (\ref{CRBT0}) becomes $\frac{\pi^2 d^2(N_r^2-1)}{3\lambda^2} \frac{r^2}{(R-r)^2}$, which tends to infinity.
\begin{figure}
	\centering 
	\includegraphics[height=4.7cm,width=6.5cm]{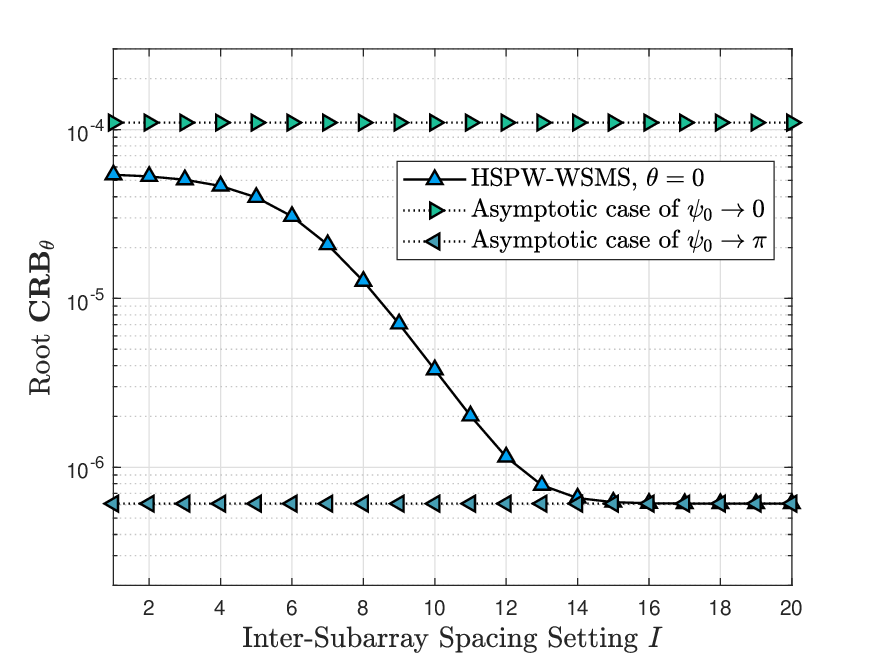}
	\caption{The angle CRB versus $I$.}\label{Asym} 
\end{figure}

The asymptotic case for HSPW-WSMS is evaluated in Fig. \ref{Asym}, where the upper and lower bounds are calculated using Eqs. \ref{WCRBR} and (\ref{WCRBT}), respectively. The parameter set considered is $\{N_r=12,\theta=0,r=10,R=50,K=2,I\in[0,20]\}$. As the value of $I$ increases, the angular span $\psi_0$ approaches $\pi$, and the CRB tends to converge to the lower bound. It should be noted that the CRB does not reach the upper bound due to the presence of a subarray size that results in $\psi_0>0$.

\begin{figure}[htbp]
	\centering
	\subfigure[Root $\textbf{CRB}_\theta$]{
		\includegraphics[width=2.62in]{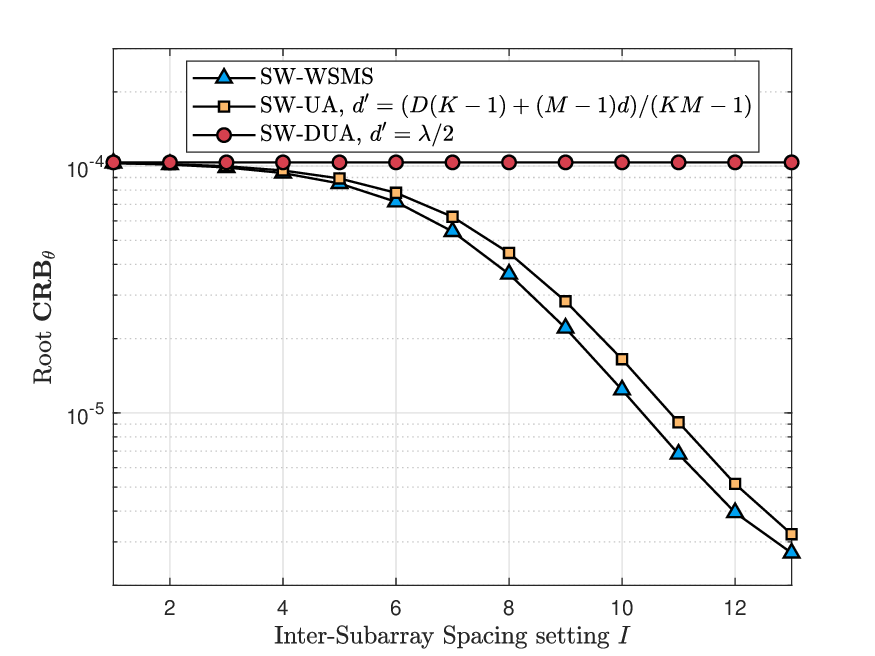}
	}
	\quad    
	\subfigure[Root $\textbf{CRB}_r$]{
		\includegraphics[width=2.62in]{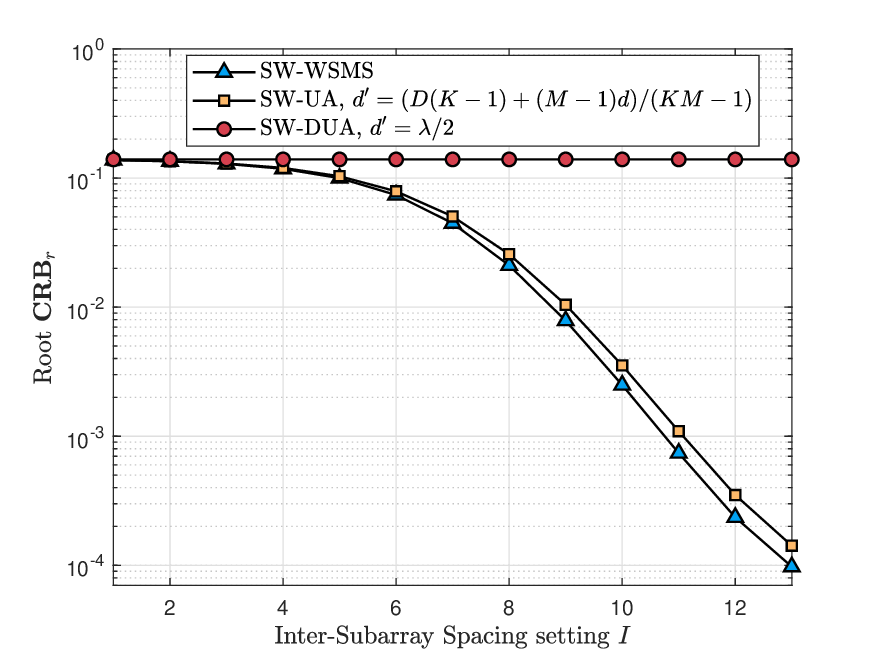}
	}
	\caption{The root CRBs of SW-WSMS, SW-UA, and SW-DUA with the same number of antennas.}
	\label{AL}
\end{figure}

Lastly, we compare the CRBs for three different array layouts: WSMS, UA, and DUA. The key similarity among them is that they all have the same number of antennas. The parameter set considered is $\{N_r=1,\theta=0,r=10,K=3,I\in[1,13]\}$, where $I=0$ is set for the SW-DUA.
As depicted in Fig. \ref{AL} (a) and (b), an increase in the parameter $I$ leads to a corresponding decrease in the CRBs of SW-WSMS and SW-UA. Furthermore, it is observed that SW-WSMS achieves a lower CRB compared to SW-UA, despite having the same array aperture and number of antennas. Particularly, the results depicted in Fig. \ref{AL} (a) are consistent with \textbf{Proposition} \ref{UA}.
\section{Conclusions}\label{Con}
Overall, this paper presents the derivation of closed-form angle and range CRBs for the SW-WSMS and HSPW-WSMS in bi-static systems.
By analyzing the closed-form CRBs for the SW-WSMS, it is observed that the CRBs can be characterized by the angular spans $\{\psi_0,\Delta_\psi\}$.  
Furthermore, a comparison is made with two WSMS layouts that share the same angular spans but differ in the number of subarrays. It is concluded that they have the same normalized CRBs, and the difference in their CRBs depends on the received signal-to-noise ratio, which is linear with the number of subarrays.
The paper also explores the CRBs of the UA by leveraging the subarray structure. It is theoretically demonstrated that the UA's CRBs are larger than the WSMS's when $\theta=0$. We also derive the closed-form CRBs for HSPW-WSMS, and its components can be considered as decomposed from the parameters of the CRBs for SW-WSMS. Additionally, asymptotic bounds are derived for the cases where $\psi_0$ tends to 0 and $\psi_0$ tends to $\pi$ when $\theta=0$.
Finally,
several simulations are conducted, yielding important results, including 1) the CRBs for WS-WSMS and HSPW-WSMS decrease as both the inter-subarray spacing and the number of subarrays increase, 2) the angle and range CRBs exhibit opposite trends with respect to changes in $r$, 3) the PW-WSMS model shows a significant error when approximating the SW-WSMS, particularly for large array apertures, 4) a model error exists between HSPW-WSMS and SW-WSMS when the array aperture is small, but this error becomes negligible as the array aperture increases, 5) the distance between the TX and the RX significantly influences the estimation of the angle. As $r$ approaches $R$, the angle CRB experiences a sharp decrease, and 6), the SW-WSMS demonstrates a slightly lower CRB than the SW-UA when considering the same number of antennas and array aperture.
 
  For future research, the utilization of HSPW-WSMS in mmWave/THz ISAC holds great promise due to its robust sensing capabilities. Several studies on wireless communications with HSPW-WSMS have already been conducted, further supporting its potential in practical applications. The CRB serves as an index of sensing performance in the ISAC system, and therefore, optimization of the precoder/combiner involved in the CRB expression is meaningful by considering both the CRB performance and communication performance. This motivates the design of the number of subarrays and inter-subarray spacing for joint optimization of communication and sensing performances.
\begin{appendices}
\section{Derivation of $\left\{ {\mathcal{S}}_{\theta^2}, {\mathcal{S}}_{\theta}, {\mathcal{S}}_{r^2}, {\mathcal{S}}_{r}, {\mathcal{S}}_{\theta r}\right\}$} \label{XL}
Here, we will derive $\mathcal{S}_{\theta^2}$ in detail. The derivation of other functions can be done in a similar manner. First, we have
 \begin{equation}
 \begin{aligned}
\mathcal{S}_{\theta^2}=&\sum_{k=-\frac{K-1}{2}}^{\frac{K-1}{2}}\sum_{m=-\frac{M-1}{2}}^{\frac{M-1}{2}}\frac{(kD+md)^2}{{r^2-2(kD+md)r\sin\theta+(kD+md)^2}} \\
=& \sum_{k=-\frac{K-1}{2}}^{\frac{K-1}{2}} \sum_{m^\prime=\frac{kD}{d}-\frac{M-1}{2}}^{m^\prime=\frac{kD}{d}+\frac{M-1}{2}}\frac{\left(m^\prime\Delta_d\right)^2}{{1-2m^\prime\Delta_d\sin\theta+\left(m^\prime\Delta_d\right)^2  }}
\\ \overset{(a)}{\approx}&\sum_{k=-\frac{K-1}{2}}^{\frac{K-1}{2}}\frac{1}{\Delta_d} \int_{\left(\frac{kD}{d}-\frac{M}{2}\right)\Delta_{d}}^{\left(\frac{kD}{d}+\frac{M}{2}\right)\Delta_{d}} \frac{x^2}{1-2x\sin\theta+x^2} {\rm d}x,
 \end{aligned}
 \end{equation}
where $(a)$ holds for the midpoint Riemann sum rule.
 Denoted by $F_{\theta^2}(x)\triangleq \int \frac{x^2}{1-2x\sin\theta+x^2} {\rm d}x$ such that $F_{\theta^2}(k\Delta_{D}+\frac{M}{2} \Delta_{d})-F_{\theta^2}(k\Delta_{D}-\frac{M}{2} \Delta_{d})= \int_{k\Delta_{D}-\frac{M}{2} \Delta_{d}}^{ k\Delta_{D}+\frac{M}{2} \Delta_{d}} \frac{x^2}{1-2x\sin\theta+x^2} {\rm d}x$. Then, we can obtain
 \begin{equation}
 	\begin{aligned}
  \mathcal{S}_{\theta^2}&=\frac{1}{\Delta_d\Delta_D}	\sum_{k=-\frac{K-1}{2}}^{\frac{K-1}{2}} \left( F_{\theta^2}\left(k\Delta_{D}+\frac{M}{2} \Delta_{d}\right) \right. \\ & \ \  \ \ \ \  \ \ \ \  \ \ \ \  \ \ \ \  \ \ \ \  \ \ \ \left.  -F_{\theta^2}\left(k\Delta_{D}-\frac{M}{2} \Delta_{d}\right)\right)\Delta_{D} \\
 &{\approx}\frac{1}{\Delta_d\Delta_D}\int_{-\frac{K}{2}\Delta_{D}+\frac{M}{2} \Delta_{d}}^{\frac{K}{2}\Delta_{D}+\frac{M}{2} \Delta_{d}} F_{\theta^2}(x) {\rm d}x\\& \ \ \ \  \ \ \ \  \  -\frac{1}{\Delta_d\Delta_D}\int_{-\frac{K}{2}\Delta_{D}-\frac{M}{2} \Delta_{d}}^{\frac{K}{2}\Delta_{D}-\frac{M}{2} \Delta_{d}} F_{\theta^2}(x) {\rm d}x
 \\ &=
\frac{1}{\Delta_d\Delta_D}\int_{\frac{K}{2}\Delta_{D}-\frac{M}{2} \Delta_{d}}^{\frac{K}{2}\Delta_{D}+\frac{M}{2} \Delta_{d}} F_{\theta^2}(x) {\rm d}x\\ & \ \ \ \  \ \ \ \  \  -\frac{1}{\Delta_d\Delta_D}\int_{-\frac{K}{2}\Delta_{D}-\frac{M}{2} \Delta_{d}}^{-\frac{K}{2}\Delta_{D}+\frac{M}{2} \Delta_{d}} F_{\theta^2}(x) {\rm d}x\\
&\triangleq \frac{1}{\Delta_d\Delta_D}\left( G_{\theta^2}(x_4)-G_{\theta^2}(x_3)-G_{\theta^2}(x_2)+G_{\theta^2}(x_1)\right),
 	\end{aligned}
 \end{equation}
 where $G_{\theta^2}(x)$ is the indefinite integral of $F_{\theta^2}(x)$, $x_1\triangleq-\frac{K}{2}\Delta_{D}-\frac{M}{2} \Delta_{d}$, $x_2\triangleq-\frac{K}{2}\Delta_{D}+\frac{M}{2} \Delta_{d}$, $x_3\triangleq\frac{K}{2}\Delta_{D}-\frac{M}{2} \Delta_{d}$, and $x_4\triangleq\frac{K}{2}\Delta_{D}+\frac{M}{2} \Delta_{d}$.
 Furthermore, $F_{\theta^2}$ and $G_{\theta^2}$ can be derived according to Eqs. (\ref{int1})-(\ref{int3}) in Appendix \ref{integral}. Particularly, $		\frac{G_{\theta^2}(x)}{\Delta_{D}\Delta_{d}}$ is given by
 \begin{equation}\label{APC}
 	\begin{aligned}
 		\frac{G_{\theta^2}(x)}{\Delta_{D}\Delta_{d}}=&\frac{x^2}{2\Delta_{D}\Delta_{d}}+\frac{x\sin\theta}{\Delta_{D}\Delta_{d}}\ln\left|\nu_1\right| -\frac{2x\sin\theta}{\Delta_{D}\Delta_{d}}\\ &-\frac{\sin^2\theta}{\Delta_{D}\Delta_{d}}\ln\left|\nu_1\right|+\frac{2\cos\theta\sin\theta}{\Delta_{D}\Delta_{d}}\tan^{-1}\left(\nu_2\right)\\ &-\frac{\cos(2\theta)}{\Delta_{D}\Delta_{d}}  \nu_2\tan^{-1}\left(\nu_2\right) +\frac{\cos(2\theta)}{2\Delta_{D}\Delta_{d}} \ln \left|\nu_2^2+1\right|,
 	\end{aligned}
 \end{equation}
where $\nu_1\triangleq 1-2 x\sin\theta+x^2$ and $\nu_2\triangleq \frac{x}{\cos\theta}-\tan\theta$ are defined for clarity.
  
 Similar to the derivation of $\mathcal{S}_{\theta^2}$, $\left\{ {\mathcal{S}}_{\theta}, {\mathcal{S}}_{r^2}, {\mathcal{S}}_{r}, {\mathcal{S}}_{\theta r}\right\}$ are derived as follows.
\begin{equation}
	\begin{aligned}
		\mathcal{S}_{\theta}&{\approx} \sum_{k=-\frac{K-1}{2}}^{\frac{K-1}{2}}\frac{1}{\Delta_d} \int_{\left(\frac{kD}{d}-\frac{M}{2}\right)\Delta_{d}}^{\left(\frac{kD}{d}+\frac{M}{2}\right)\Delta_{d}}{\sqrt{x^2-2x\sin\theta+1}} {\rm d}x \\& = \frac{1}{\Delta_D\Delta_d}\int_{x_3}^{x_4} \left(
		\sqrt{\nu_1}+\sin\theta \tanh ^{-1}\left(\frac{x-\sin\theta}{\sqrt{\nu_1}}\right)\right){\rm d} x\\&\ \ -\frac{1}{\Delta_D\Delta_d}\int_{x_1}^{x_2} \left(
		\sqrt{\nu_1}+\sin\theta \tanh ^{-1}\left(\frac{x-\sin\theta}{\sqrt{\nu_1}}\right) \right){\rm d} x\\
	&	=\frac{1}{\Delta_d\Delta_D}\left( G_{\theta}(x_4)-G_{\theta}(x_3)-G_{\theta}(x_2)+G_{\theta}(x_1)\right),
	\end{aligned}
\end{equation}
where $G_\theta(x)$ can be derived by Eqs. (\ref{int4}) and (\ref{int5}) as
$
\frac{G_\theta(x)}{\Delta_{D}\Delta_{d}}=\frac{x-\sin\theta}{2\Delta_D\Delta_d}  \sqrt{\nu_1}+\frac{\cos^2\theta}{2\Delta_D\Delta_d} \ln \left(\sqrt{\nu_1}-\sin\theta+x\right) +\frac{\sin\theta(x-\sin\theta)}{\Delta_D\Delta_d} \tanh ^{-1}\left(\frac{x-\sin\theta}{\sqrt{\nu_1}}\right)-\frac{\sin\theta}{\Delta_D\Delta_d}\sqrt{\nu_1}.$.
\begin{equation}
	\begin{aligned}
		\mathcal{S}_r
	 {\approx}& \sin\theta \mathcal{S}_\theta -\sum_{k=-\frac{K-1}{2}}^{\frac{K-1}{2}}\frac{1}{\Delta_d} \int_{\left(\frac{kD}{d}-\frac{M}{2}\right)\Delta_{d}}^{\left(\frac{kD}{d}+\frac{M}{2}\right)\Delta_{d}}
		\frac{1}{\sqrt{1-2x\sin\theta+x^2}}{\rm d}x 		\\ =& 
		 \sin\theta \mathcal{S}_\theta -\frac{1}{\Delta_{d}\Delta_{D}}\int_{x_3}^{x_4}\ln\left| \sqrt{\nu_1}+x-\sin\theta\right|{\rm d}x\\&   + \frac{1}{\Delta_{d}\Delta_{D}}\int_{x_1}^{x_2}\ln\left| \sqrt{\nu_1}+x-\sin\theta\right|{\rm d}x  	\\ =& 
			 \sin\theta \mathcal{S}_\theta-\frac{1}{\Delta_d\Delta_D}\left( G_{r}(x_4)-G_{r}(x_3)-G_{r}(x_2)+G_{r}(x_1)\right),
	\end{aligned}
\end{equation}
 where $\frac{G_r(x)}{\Delta_{D}\Delta_{d}}= \frac{1}{\Delta_{d}\Delta_{D}} (x-\sin\theta)   \ln\left| \sqrt{\nu_1}+x-\sin\theta\right|-\frac{1}{\Delta_{d}\Delta_{D}}\sqrt{\nu_1}$, which can be derived by Eqn. (\ref{int6}).
 \begin{equation}
 	\begin{aligned} 
 		\mathcal{S}_{r^2}=KM-\cos^2\theta \mathcal{S}_{\theta^2}.
 	\end{aligned}
 \end{equation}
 \begin{equation}
 	\begin{aligned}
 	&	\mathcal{S}_{\theta r}
 		 {\approx}\sin\theta \mathcal{S}_{\theta^2}-  \sum_{k=-\frac{K-1}{2}}^{\frac{K-1}{2}}\frac{1}{\Delta_d} \int_{\left(\frac{kD}{d}-\frac{M}{2}\right)\Delta_{d}}^{\left(\frac{kD}{d}+\frac{M}{2}\right)\Delta_{d}}
 		\frac{x}{{1-2x\sin\theta+x^2}}{\rm d}x \\ 
 		&=\sin\theta \mathcal{S}_{\theta^2}-\frac{1}{\Delta_{D}\Delta_{d}}\int_{x_3}^{x_4} \tan\theta \tan^{-1}\left(\nu_2\right)+\frac{1}{2}\ln\left|\nu_1\right| {\rm d}x  \\& \ \ \ \  +\frac{1}{\Delta_{D}\Delta_{d}}\int_{x_1}^{x_2} \tan\theta \tan^{-1}\left(\nu_2\right)+\frac{1}{2}\ln\left|\nu_1\right| {\rm d}x\\
 		&=\sin\theta\mathcal{S}_{\theta^2}- \frac{1}{\Delta_d\Delta_D}\left( G_{\theta r}(x_4)-G_{\theta r}(x_3)-G_{\theta r}(x_2)+G_{\theta r}(x_1)\right),
 	\end{aligned}
 \end{equation}
where $
\frac{G_{\theta r}(x)}{\Delta_{D}\Delta_{d}}=\frac{\sin\theta}{\Delta_{D}\Delta_{d}} \left( \nu_2 \tan^{-1}\left(\nu_2\right)-\frac{1}{2}\ln \left|\nu_2^2+1\right|\right)+\frac{x}{2\Delta_{D}\Delta_{d}}\ln\left|\nu_1\right| -\frac{ x}{\Delta_{D}\Delta_{d}}-\frac{\sin\theta}{2\Delta_{D}\Delta_{d}}\ln\left|\nu_1\right| +\frac{ \cos\theta}{\Delta_{D}\Delta_{d}}\tan^{-1}\left(\nu_2\right)$.
 
	\section{Derivation of $\left\{\widetilde{\mathcal{S}}_{\theta^2},\widetilde{\mathcal{S}}_{\theta},\widetilde{\mathcal{S}}_{r^2},\widetilde{\mathcal{S}}_{r},\widetilde{\mathcal{S}}_{\theta r}\right\}$} \label{WS}
 Here, we give the closed-form expressions of $\left\{\widetilde{\mathcal{S}}_{\theta^2},\widetilde{\mathcal{S}}_{\theta},\widetilde{\mathcal{S}}_{r^2},\widetilde{\mathcal{S}}_{r},\widetilde{\mathcal{S}}_{\theta r}\right\}$ as follows. For clarity, we define $\kappa_1\triangleq 1-K\Delta_D\sin\theta+K^2\Delta_{D}^2/4$ and $\kappa_2\triangleq 1+K\Delta_D\sin\theta+K^2\Delta_{D}^2/4$. 
 	\begin{equation}\label{WST2}
 		\begin{aligned}
 	 		\widetilde{\mathcal{S}}_{\theta^2} =&\frac{1}{\Delta_{D}} \int_{-\frac{K\Delta_{D}}{2}}^{\frac{K\Delta_{D}}{2}} \frac{x^2}{1-2x\sin\theta+x^2}{\rm d}x \\
 	 		=& K+\frac{\sin\theta}{\Delta_{D}} \ln \left| \frac{\kappa_1}{\kappa_2}\right|  -\frac{\cos(2\theta)}{\Delta_{D}\cos(\theta)}\tan^{-1}\left(\frac{K\Delta_{D}}{2\cos\theta}-\tan\theta\right)\\ &+\frac{\cos(2\theta)}{\Delta_{D}\cos(\theta)}\tan^{-1}\left(\frac{-K\Delta_{D}}{2\cos\theta} -\tan\theta\right).
 		\end{aligned}
 	\end{equation}
 \begin{equation}
 	\begin{aligned}
 	\widetilde{\mathcal{S}}_\theta
 	=& \frac{1}{\Delta_{D}} \int_{-\frac{K\Delta_{D}}{2}}^{\frac{K\Delta_{D}}{2}} \frac{x}{\sqrt{1-2x\sin\theta+x^2}}{\rm d}x\\
 	 =& \frac{ 	\sin\theta}{\Delta_{D}}
 \tanh ^{-1}\left(\frac{K\Delta_{D}/2-\sin\theta}{\sqrt{\kappa_1}}\right)-  \frac{ 		\sqrt{\kappa_2}}{\Delta_{D}}  +  	 \frac{\sqrt{\kappa_1} }{\Delta_{D}}\\&
  -\frac{\sin\theta}{\Delta_{D}} \tanh ^{-1}\left(\frac{-K\Delta_{D}/2-\sin\theta}{\sqrt{\kappa_2}}\right).
 	\end{aligned}
 \end{equation}
\begin{equation}
	\begin{aligned}
		\widetilde{\mathcal{S}}_{r^2}=& K-\sum_{k=1}^{K}\frac{n_k^2\cos^2\theta}{{r^2-2n_kr\sin\theta+n_k^2}}   \\ 
		=& K-  {\cos^2\theta}\widetilde{\mathcal{S}}_{\theta^2}.
	\end{aligned}
\end{equation}
\begin{equation}
	\begin{aligned}
		\widetilde{\mathcal{S}}_{r}=&\sin\theta \widetilde{\mathcal{S}}_\theta-
		\sum_{k=1}^{K}\frac{ r}{\sqrt{r^2-2n_kr\sin\theta+n_k^2}} \\
		=& \sin\theta \widetilde{\mathcal{S}}_\theta-\frac{1}{\Delta_D}\ln\left| \frac{	  \sqrt{\kappa_1}+K\Delta_{D}/2-\sin\theta}{\sqrt{\kappa_2}-K\Delta_{D}/2-\sin\theta}\right|.
	\end{aligned}
\end{equation}
\begin{equation}
	\begin{aligned}
	\widetilde{\mathcal{S}}_{\theta r}=&\sin\theta \widetilde{\mathcal{S}}_{\theta^2}-\sum_{k=1}^{K}\frac{rn_k }{{r^2-2n_kr\sin\theta+n_k^2}}\\
	= &\sin\theta \widetilde{\mathcal{S}}_{\theta^2}-\frac{\tan\theta}{\Delta_{D}} \tan^{-1}\left(\frac{K\Delta_{D}}{2\cos\theta}-\tan\theta\right)\\&+ \frac{\tan\theta}{\Delta_{D}} \tan^{-1}\left(-\frac{K\Delta_{D}}{2\cos\theta}-\tan\theta\right)
	-\frac{1}{2\Delta_{D}}\ln \left| \frac{\kappa_1 }{\kappa_2}  \right|.
	\end{aligned}
\end{equation}
	\section{Derivation of Some integrals used in this paper } \label{integral}
	Here, some useful intergrals are derived to support Appendix. \ref{XL} and \ref{WS}.  $\nu_1\triangleq 1-2 x\sin\theta+x^2$ and $\nu_2\triangleq \frac{x}{\cos\theta}-\tan\theta$ are defined in derivation results for clarity.
\begin{equation}\label{int1}
	\begin{aligned}
		\int \frac{x^2}{\nu_1}{\rm d}x&=\int \frac{\nu_1}{\nu_1}{\rm d}x+\int \frac{2x\sin\theta-1}{\nu_1}{\rm d}x\\&
		\overset{(b)}{=}x+\sin\theta \ln \left|\nu_1 \right| -\frac{\cos(2\theta)}{\cos(\theta)}\tan^{-1}\left(\nu_2\right)+Const,
	\end{aligned}
\end{equation}
where $(b)$ holds due to $\int \frac{m x+n}{a x^2+b x+c} d x=\frac{m}{2 a} \ln \left|a x^2+b x+c\right|+\frac{2 a n-b m}{a \sqrt{4 a c-b^2}} \tan^{-1} \frac{2 a x+b}{\sqrt{4 a c-b^2}}$ for $4ac-b^2\geq 0$.
\begin{equation}\label{int2}
	\begin{aligned} 
	&\int\ln\left|\nu_1\right|{\rm d}x	=x\ln\left|\nu_1\right|-\int \frac{2x^2-2\sin\theta x}{ \nu_1 } {\rm d}x\\& =x\ln\left|\nu_1\right|-\int \frac{2x^2-4\sin\theta x+2+2\sin\theta x-2}{\nu_1} {\rm d}x\\ & {=} x\ln\left|\nu_1\right| -2x-\sin\theta\ln\left|\nu_1\right|+2\cos\theta\tan^{-1}\left(\nu_2\right)+Const.
	\end{aligned}
\end{equation} 
\begin{equation}\label{int3}
	\int \tan^{-1}\left(\nu_2\right){\rm d}x=   \cos\theta \left(\nu_2\tan^{-1}\left(\nu_2\right)-\frac{1}{2}\ln \left|\nu_2^2+1\right|\right)
\end{equation}
\begin{equation}
	\begin{aligned}
		\int\frac{x}{\sqrt{\nu_1}} {\rm d}x =
		\sqrt{\nu_1}+\sin\theta \tanh ^{-1}\left(\frac{x-\sin\theta}{\sqrt{\nu_1}}\right)+Const.
	\end{aligned}
\end{equation}
\begin{equation}\label{int4}
	\begin{aligned}
		 \int \sqrt{\nu_1} d x= \frac{x-\sin\theta}{2}  \sqrt{\nu_1}+\frac{\cos^2\theta}{2} \ln \left(\sqrt{\nu_1}-\sin\theta+x\right)
	\end{aligned}
\end{equation}
\begin{equation}\label{int5}
	\begin{aligned}
	&	\int \tanh ^{-1}\left(\frac{x-\sin\theta}{\sqrt{\nu_1}}\right) {\rm d}x=\\& \ \ \ (x-\sin\theta) \tanh ^{-1}\left(\frac{x-\sin\theta}{\sqrt{\nu_1}}\right)-\sqrt{\nu_1}
		+Const,
	\end{aligned}
\end{equation}
\begin{equation}\label{int6}
	\begin{aligned}
	&	\int \ln\left| \sqrt{\nu_1}+x-\sin\theta\right|{\rm d}x= \\ & \ \ \ \ \ \ (x-\sin\theta)   \ln\left| \sqrt{\nu_1}+x-\sin\theta\right|-\sqrt{\nu_1}+Const.
	\end{aligned}
\end{equation}
\end{appendices}
\bibliographystyle{IEEEtran}
\bibliography{reference.bib}

\vspace{12pt}

\end{document}